\documentclass[a4paper,11pt]{article}
\pdfoutput=1 
\usepackage{jcappub}

\usepackage{amsmath}
\usepackage{graphicx}
\usepackage{latexsym}
\usepackage{xspace}
\usepackage[most]{tcolorbox}
\usepackage{color}
\usepackage{hyperref} 
\usepackage{bm}
\usepackage{relsize}
\usepackage{tabularx}
\usepackage{multirow}
\usepackage{amssymb}
\usepackage{dsfont}
\usepackage[]{mdframed}
\usepackage{mathtools}
\usepackage{slashed}
\usepackage{fullpage}
\usepackage{xcolor}
\usepackage{cancel}

\makeatletter
\gdef\@fpheader{}
\g@addto@macro\bfseries{\boldmath}
\makeatother

\newcommand{\Ji}[1]{{\color{red} #1} }

\newcommand{\ie}{{i.e.~}}

\newcommand{\etc}{{etc.\xspace}}

\newcommand{\cs}{c_{_\mathrm{S}}}

\newcommand{\beaq}{\begin{equation}\begin{aligned}}
\newcommand{\eeaq}{\end{aligned}\end{equation}}

\newlength{\wsingfig}
\newlength{\wdblefig}
\newlength{\wquadfig}
\newlength{\wtriplefig}
\newcommand{\Eq}[1]{Eq.~(\ref{#1})}
\newcommand{\Eqs}[1]{Eqs.~(\ref{#1})}

\newcommand{\Sec}[1]{Section~\ref{#1}}

\newcommand{\App}[1]{Appendix~\ref{#1}}

\newcommand{\etad}[1]{{#1^{\prime}}}
\newcommand{\etadd}[1]{{#1^{\prime\prime}}}

\newcommand{\gammaB}{\gamma}
\newcommand{\DB}{D}
\newcommand{\DeltaB}{\Delta}

\newcommand{\cB}{{\mathcal B}}
\newcommand{\cE}{{\mathcal E}}

\newcommand{\cJ}{{\mathcal J}}

\newcommand{\cL}{{\mathcal L}}
\newcommand{\cH}{{\mathcal H}}
\newcommand{\cM}{{\mathcal M}}

\newcommand{\cR}{{\mathcal R}}
\newcommand{\cO}{{\mathcal O}}

\newcommand{\cT}{{\mathcal T}}
\newcommand{\cV}{{\mathcal V}}

\newcommand{\cC}{{\mathcal C}}

\newcommand{\cS}{{\mathcal S}}

\newcommand{\cI}{{\mathcal I}}

\newcommand{\SO}{\mathrm{SO}}

\newcommand{\be}{\begin{equation}}
\newcommand{\ee}{\end{equation}}
\newcommand{\beq}{\begin{eqnarray}}
\newcommand{\eeq}{\end{eqnarray}}
\newcommand{\bes}{\begin{eqnarray}}
\newcommand{\ees}{\end{eqnarray}}

\newcommand{\mat} [2] {\left ( \begin{array}{#1}#2\end{array} \right ) }

\newcommand{\so}{{\mathfrak{so}}}

\definecolor{purple}{rgb}{0.57, 0.36, 0.75}
\definecolor{bittersweet}{rgb}{1.0, 0.44, 0.37}

\def\rd{\textrm{d}}

\def\beq{\begin{eqnarray}}
\def\eeq{\end{eqnarray}}
\def\be{\begin{equation}}
\def\ee{\end{equation}}

\def\r2A{z}

\def\nn{\nonumber}

\numberwithin{equation}{section}

\title{Symmetries of cosmological perturbations: \\ The residual low-multipole ambiguity}
\author[a,b]{Jibril Ben Achour,}
\author[b]{Etera R. Livine,}
\author[c]{Vincent Vennin}

\affiliation[a]{Arnold Sommerfeld Center for Theoretical Physics, Munich, Germany}
\affiliation[b]{Univ Lyon, CNRS, ENS de Lyon, Laboratoire de Physique (LPENSL), Lyon, France}
\affiliation[c]{Laboratoire de Physique de l'Ecole Normale Sup\'erieure, ENS, CNRS, Universit\'e PSL, Sorbonne Universit\'e, Universit\'e Paris Cit\'e, 75005 Paris, France}

\emailAdd{jibril.ben-achour@ens-lyon.fr}
\emailAdd{etera.livine@ens-lyon.fr}
\emailAdd{vincent.vennin@phys.ens.fr}

\date{today}

\begin{document}
\sloppy

\abstract{In cosmology, long-wavelength modes are related to large-gauge transformations (LGT), \ie changes of coordinates that modify the physical geometry of the cosmological patch. These LGTs stand as bona-fide symmetries of cosmological perturbation theory with various applications, from consistency relations constraining cosmological correlators to non-linear conservation laws in the separate-universe approach. In this work, we revisit LGTs and derive two new results. First, we show that the global symmetries already identified in the literature can be extended to local infinite-dimensional symmetries. The associated generators depend on arbitrary functions of time, and generate low-multipole modes that modify the mean curvature energy and the angular momentum of the patch, demonstrating their physical nature. We propose to interpret these low-multipole soft modes as a new \textit{cosmological-frame ambiguity} that needs to be fixed prior to evaluating cosmological observables. Second, we demonstrate that the adiabatic cosmological perturbations generated by LGTs deform but preserve all the explicit and hidden Killing symmetries of the background geometry. As such, long-wavelength modes stand as a concrete example of algebraically-special cosmological perturbations of Petrov-type O, and inherit the conformal group as isometries and a set of four deformed Killing-Yano tensors and their associated Killing tensors. This opens the possibility to study their effect on cosmological observables in a fully analytic manner. }


\maketitle

\newpage

\section{Introduction}

The observations of the temperature fluctuations of the cosmic microwave background (CMB) in the last decades have produced stringent constraints on the state of the early universe at recombination, \ie at redshift $z\sim 1100$, constraining to a great accuracy the parameters of the $\Lambda$CDM model and the landscape of inflationary models \cite{Planck:2018jri, Martin:2013nzq, Chowdhury:2019otk, Martin:2024qnn}. In parallel, the large-scale surveys such as the ongoing DESI and EUCLID missions \cite{DESI:2024mwx, Euclid:2024yrr}, enable us to test the cosmic evolution and the formation of large-scale structure through a precise observation of the distribution and the shapes of millions of galaxies in a region extending up to redshifts $z \sim 2$. This allows us to test the composition and the dynamics of the universe during the transition between the matter-dominated and the dark-energy dominated epochs at which the observed accelerated expansion started, therefore providing an unprecedented window on the nature of the dark energy. With this vast amount of data incoming, refined techniques are required to ensure a robust interpretation of observations free from theoretical bias.

As usual in gravity, care must be taken when choosing an adapted coordinate system to compute cosmological observables and compare them to observations. The subtle role played by the gauge transformations, \ie the changes of space-time coordinates, manifests at different levels in this process. First, even after having fixed a gauge to describe the geometry and the matter fields, one can identify residual changes of coordinates that preserve the gauge but that have non-trivial effects on the description of the cosmological patch. This is so because in the presence of a boundary, one has to distinguish the changes of coordinates that leave it invariant from the ones that modify the boundary. This distinction is crucial inasmuch as gauge-invariant observables (energy, angular momentum, higher multipoles, \etc) of a finite region of spacetime are quasi-local, \ie they take the form of integrals over the boundary of the spacetime region of interest. Therefore, ``small'' gauge transformations that are associated to vanishing boundary charges do not modify the gravitational system, while ``large'' gauge transformations (LGTs) that are associated with non-vanishing boundary charges do. In general, they correspond to flux-balance laws encoding how observables evolve in the presence of a flux of radiative degrees of freedom through the boundary and reduce to conserved charges only in the absence of flux. Their expressions can be computed using the covariant phase-space formalism \cite{Compere:2018aar, Harlow:2019yfa, Freidel:2020xyx, Ciambelli:2022vot}, hence LGTs can be defined only w.r.t. to a given choice of gauge and boundary. As a consequence, LGTs stand as changes of coordinates that are upgraded to physical symmetries of the gravitational system by the presence of the boundary. These soft symmetries are of utmost importance when characterizing the infrared regime of the system. Indeed, being a physical symmetry, a LGT maps gauge-inequivalent solutions of the linearized Einstein equations onto another, acting as a solution-generating map w.r.t. the solution space. Concretely, two solutions differ by the presence of a soft mode, \ie a long wavelength mode (LWM) that modifies the geometrical properties of the cosmological patch. Indeed, by construction, adding or removing a LWM to the bulk metric is not harmless as the associated change of coordinates affects the boundary properties of the cosmological patch, and thus the geometrical model one is using to capture the large-scale geometry of the observable universe.

The role of LWMs for early and late cosmology was recognized a long time ago. As first shown by Weinberg \cite{Weinberg:2003sw}, the solution-generating point of view on LGTs can be used to construct a large family of cosmological-perturbation profiles that are conserved on super-Hubble scales. These are precisely the adiabatic modes that allow one to connect the curvature and tensor perturbations generated during inflation when they exit the Hubble radius, with the observed large-scale fluctuations in the CMB. They also play a crucial role in the separate-universe approach~\cite{Artigas:2021zdk, Tanaka:2021dww, Cruces:2022dom, Jackson:2023obv, Artigas:2024ajh, Tanaka:2024mzw, Briaud:2025ayt}. In turn, these soft symmetries translate into key consistency relations for cosmological correlators in the squeezed limit. The most well-known one is the Maldacena relation constraining the non-Gaussianities in single-field inflation \cite{Maldacena:2002vr}. Generalizations for higher inflationary correlators were worked out in \cite{Bartolo:2001rt, Creminelli:2004yq, Cheung:2007sv, Creminelli:2011sq, Creminelli:2012ed, Hinterbichler:2012nm, Goldberger:2013rsa, Hinterbichler:2013dpa, Pajer:2013ana, Mirbabayi:2014zpa, Mirbabayi:2016xvc, Hinterbichler:2016pzn, Finelli:2017fml, Akhshik:2015rwa, Hui:2018cag, Grall:2020ibl} while similar consistency relations constraining the correlation functions of the density contrast have been obtained in \cite{Kehagias:2013yd, Creminelli:2013nua, Creminelli:2013mca, Goldstein:2022hgr}, extending these constraints to the growth of large-scale structures. In parallel, important efforts have been devoted to clarifying how LWMs affect various cosmological observables, from the CMB low multipoles \cite{Zibin:2008fe, Baumgartner:2020llb} to the observed galaxy bi-spectrum \cite{Castorina:2021xzs} or the luminosity-distance relation \cite{Kolb:2004am, Barausse:2005nf, Biern:2016kys} to mention a few. See also \cite{Matarrese:2020why, Bartolo:2022wqq, Mitsou:2022fvy, Mitsou:2023wes, Blachier:2023ooc} for related discussions on this long standing debate. In view of the key role played by the LWMs in early and late-time cosmology, it appears crucial to have a complete characterization of these soft cosmological symmetries. 

In the cosmological context, LGTs and their associated LWMs have been mostly studied in the Newtonian gauge \cite{Weinberg:2003sw, Creminelli:2013mca, Pajer:2017hmb} and the uniform-density gauge (or $\zeta$ gauge) \cite{Creminelli:2011sq, Creminelli:2012ed, Hinterbichler:2012nm}. Surprisingly, it was shown that irrespectively of the gauge choice, one can identify an infinite set of soft symmetries that are nevertheless global, \ie the generators of the LGTs do not involve any free functions of the spacetime coordinates. It implies that the profiles of the generated LWMs are completely fixed up to free parameters. The most well-known profiles are the first and second Weinberg modes, which are constant or linear curvature perturbations built respectively from a constant dilatation and a special conformal transformation acting on the spatial hypersurface \cite{Weinberg:2003sw}. But a rich and infinite family of scalar, vector and tensor adiabatic modes can be obtained along the same line \cite{Pajer:2017hmb}. This outcome strongly contrasts with the status of LGTs of asymptotically flat spacetimes (AFS), which are best described within the Bondi gauge. 

In this asymptotically-flat context, the LGTs are infinite dimensional and local. They form the well-known Bondi-Metzner-Sachs (BMS) group which extends the Poincar\'e isometries of Minkowski space-time by a set of free angle-dependent time-translations and rotations respectively called supertranslations and superrotations \cite{Bondi:1960jsa, Bondi:1962rkt, Bondi:1962px}. This result holds both at null infinity $\cI^{+}$ \cite{Barnich:2009se, Barnich:2010eb, Barnich:2011mi, Barnich:2012nkq, Barnich:2016lyg, Barnich:2019vzx} and at spatial infinity $i_0$ \cite{Henneaux:2018cst, Henneaux:2019yax}. See \cite{Fiorucci:2021pha} for a review, and \cite{Campiglia:2020qvc, Flanagan:2015pxa, Compere:2020lrt, Compere:2018ylh, Freidel:2021fxf, Geiller:2022vto, Geiller:2024amx} for extensions of the BMS symmetries. A direct and crucial consequence of the local nature of the BMS transformations is that they generate soft modes whose (angular) profiles are unconstrained, introducing an ambiguity in the notion of AFS. These unconstrained modes give rise to the so-called memory observables, a major target of gravitational-wave astronomy for the decade to come. To see this, consider a supertranslation of the null Bondi time $u$ which reads $u \rightarrow u - \alpha (\theta, \varphi)$ where $(\theta, \varphi)$ are the coordinates of the 2-sphere. Analyzing the gravitational waves emitted by a binary black-hole merger event w.r.t. to a given AFS requires to fix a choice of the free function $\alpha (\theta, \varphi)$. Since it corresponds to a LGT, it is associated to a boundary charge whose value can be used to control this ambiguity and to characterize w.r.t. to which AFS geometry this gravitational-wave event is considered. This is known as the \textit{BMS frame ambiguity} and its role in the analysis of gravitational-wave templates has been recognized as crucial~\cite{Mitman:2021xkq, Mitman:2022kwt, Iozzo:2021vnq, MaganaZertuche:2021syq, Compere:2019gft, Blanchet:2020ngx, Blanchet:2023pce}.  See \cite{Mitman:2024uss} for a pedagogical review on this point.

From this discussion, a natural question is whether or not a similar frame ambiguity exists in the cosmological context. In other words, could there be local soft symmetries of the cosmological perturbations that generate LWMs with unconstrained profiles? 

The first goal of this work is to reinvestigate LGTs for linear cosmological perturbations in the Newtonian gauge, in order to address this crucial question. As we shall see, one can indeed identify extended local soft symmetries in this context. These local LGTs generate new low multipole soft modes whose time dependence is unconstrained. These modes were not considered physical in previous investigations \cite{Weinberg:2003sw, Creminelli:2013mca, Pajer:2017hmb, Creminelli:2011sq, Creminelli:2012ed, Hinterbichler:2012nm}. The main reason is that the very concept of soft modes used in cosmology, as introduced by Weinberg in \cite{Weinberg:2003sw}, is more restrictive than the notion of soft modes studied in other contexts, in particular at the interface between gravitational waves and asymptotic symmetries where the role of the boundary is crucial. Indeed, following Weinberg \cite{Weinberg:2003sw}, a soft mode generated by a LGT is said ``physical'' if it satisfies the linearized Einstein equations away from the zero-momenta limit in Fourier space, \ie $k\rightarrow 0$. In contrast, hereafter a soft mode is declared ``physical'' when the diffeomorphism that generates it modifies the boundary charges, hence modifying quasi-local observables such as the quasi-local (mean curvature) energy or the angular momentum of the patch. Besides lacking a clear geometrical interpretation, Weinberg's prescription
imposes solving more restrictive versions of the linearized Einstein equations, missing those unconstrained soft modes that precisely live in the kernel of spatial gradient operators. After having described these unconstrained cosmological soft modes, we further clarify their effects on the boundary by computing the quasi-local energy and the angular momentum of the cosmological patch. As expected, we shall show that the soft monopole and dipole modify the quasi-local mass and the angular momentum of the patch. This result is not surprising and parallels the well-known role of monopole and dipole perturbations in black-hole theory where monopole and (axial) dipole perturbations of the Schwarzschild black hole respectively shift its mass and make the hole slowly rotate.     

This first result provides a preliminary step to fully recognize the physicality of these low-multipole soft modes allowed by the linearized Einstein equations. Moreover, since their time dependence is dynamically unconstrained, it suggests the existence of a frame-ambiguity affecting the description of the cosmological patch. The next step, which goes beyond the scope of this work, will require a careful analysis of the associated flux-balance laws that can be derived from the covariant phase-space formalism. However, the analysis presented in this work already demonstrates that the linearized Einstein's equations admit free time-dependent monopole and dipole contributions that could affect non-trivially the interpretation of cosmological observables, modifying in particular the energy and angular momentum of the cosmologocial patch.

The second goal of this work is to emphasize the role of the LGTs in deriving the Killing symmetries of the cosmological Friedmann-Lema\^itre-Robertson-Walker (FLRW) patch in the presence of non-trivial LWMs. Contrary to the soft symmetries which stand as symmetries of the theory, the Killing symmetries are symmetries of the solution. Thus they are crucial to analyze the test probes on a given inhomogeneous cosmological geometry, such as the propagation of test particles or fields. We first point out that any LWM dressing the flat FLRW geometry preserves the Killing symmetries of the FLRW background. This result could be misleadingly considered as trivial since it is a direct consequence of covariance, but the key point is that the background and deformed FLRW geometries are gauge-inequivalent and thus correspond to two distinct geometries. This thus leads to the identification of a whole class of large-scale inhomogeneities that share the same properties as the FLRW background. As a direct consequence, the LWMs preserve the Petrov type O of the flat FLRW geometry. As such, these adiabatic LWMs stand as one concrete example of algebraically-special cosmological perturbations of the simplest Petrov type, paralleling the notion of algebraically-special perturbations for black holes studied in \cite{Couch:1973zc, Wald:1973wwa, Qi:1993ey, BenAchour:2025hns, BenAchour:2024skv}. Far from being of mere academic importance, this property opens a path to identifying hidden symmetries in the presence of large-scale cosmological perturbations. As a concrete example, we point out that the four Killing-Yano tensors of the FLRW background can be readily Lie-dragged to non-trivial Killing-Yano tensors for the LWMs constructed via LGTs. The interest in identifying these explicit and hidden symmetries is that they are crucial tools to analytically study key problems in cosmology, such as the effects of large-scale inhomogeneities on the propagation of photons or gravitational waves. Finally, let us mention that the notion of algebraically-special perturbations provides a rigorous way to identify which LWM can be introduced without spoiling the FLRW property. This could prove useful for the separate-universe approach and the closely related stochastic-inflation formalism.

This work is organized as follows. In \Sec{A1}, we present a brief overview of cosmological perturbation theory where we fix our notations. In \Sec{A2}, we introduce the Newtonian gauge and derive gauge-fixed equations. The generation of soft perturbations is discussed in \Sec{A3}. The identification of extended soft symmetries is detailed in \Sec{A4}. \Sec{D} is devoted to the study of the physical effects of the low soft multipoles on the quasi-local properties of the cosmological patch, namely its quasi-local energy and angular momentum. We first introduce the definition of the mean-curvature energy and spin and compute their value in the presence of monopole and dipole soft modes. In the next \Sec{B}, we study the Killing symmetries of the cosmological patch dressed with LWMs. After recalling how Killing symmetries can be Lie dragged in \Sec{B1}, we review the explicit and hidden Killing symmetries of the FLRW background in \Sec{B2}, focusing on the case of a power-law scale factor. Then, we derive expressions for both the (conformal) Killing vectors and the Killing-Yano tensors, in terms of the components of the soft symmetry generators for any inhomogeneous cosmological geometry constructed from LGT of the seed flat FLRW geometry. The last \Sec{C} is devoted to a discussion on the extended soft symmetries found in this work, and the open perspectives triggered by both the soft and Killing symmetries. Finally, the paper ends with a few appendices to which technical details are deferred.

\section{Extended soft symmetries for adiabatic cosmological perturbations}

\label{A}

In this section, we revisit and present a detailed derivation of the soft symmetries of adiabatic cosmological perturbations in the Newtonian gauge. See \cite{Pajer:2017hmb} for a systematic investigation of the soft modes in this gauge. We present two new results. First, we exhibit a class of soft symmetry generators that contain monopole and dipole contributions that are not fixed but that involve free time-dependent functions. This makes the soft symmetry an infinite-dimensional local symmetry, instead of a global symmetry. Second, we show that, as a direct consequence, in the absence of anisotropic pressure, the profiles of the scalar potentials can differ at the monopole and dipole orders. 

\subsection{Cosmological perturbation theory in a nutshell}

\label{A1}

Let us start by reviewing the kinematics and the dynamics of linearized cosmological perturbations without gauge fixing, in order to set up our notation. See \cite{Kodama:1984ziu, Mukhanov:1990me, Brandenberger:2003vk, Baumann:2009ds, Peter:2013avv, Mukhanov:2013tua, Baumann:2018muz, Poly} for pedagogical reviews. 
Consider the perturbed FLRW metric of the form
\begin{align}
\label{pertmet}
\rd s^2 = a^2 \left[ - (1+2 A) \rd \eta^2 + 2N_i \rd \eta \rd x^i + \left({\gammaB}_{ij} + h_{ij}\right) \rd x^i \rd x^j\right]
\end{align}
where $(A, N_i, h_{ij})$ are cosmological perturbations. In what follows, $\gamma_{ij}$ will refer to the euclidean three-dimensional background metric. In particular, we do not consider perturbations about the closed or open FLRW geometries but only the flat one.
Thanks to the scalar-vector-tensor (SVT) decomposition, the perturbations $(N_i, h_{ij})$ can be split into
\begin{align}
N_i & = {\DB}_i B + B_i  \\
h_{ij} & = 2C {\gammaB}_{ij} + 2 {\DB}_i {\DB}_j E + 2 {\DB}_{(i} E_{j)} + 2 E_{ij} 
\end{align}
such that the vectors $B_i$ and $E_i$ are divergenceless and that the tensor $E_{ij}$ is both divergenceless and traceless, \ie 
\be
\label{eq:divergenceless:etc}
{\DB}_i B^i ={\DB}_i E^i =0\, ,  \qquad {\DB}_i E^i{}_j =0\, , \qquad E^i{}_i =0 \, .
\ee
Here, ${\DB}_i$ denotes the covariant derivative with respect to the induced background metric ${\gammaB}_{ij}$.
Out of the ten perturbation fields $A, C, B, E, E_i,B_i, E_{ij}$, we can construct six gauge-invariant variables. Two of them correspond to the gravitational waves $E^{ij}$ and the remaining four are given by the two Bardeen potentials $(\Phi, \Psi)$ and the vector $\Phi^i$
\begin{align}
\label{GI1}
\Psi & = - C - \cH (B - \etad{E})\, , \\
\label{GI2}
\Phi & = A + \cH (B - \etad{E}) + (\etad{B} - \etadd{E})\, , \\
\label{GI3}
\Phi^i & = (E^i)' - B^i\, ,
\end{align}
where a prime denotes derivation with respect to conformal time $\eta$ and $\mathcal{H}=\etad{a}/a$. Let us now turn to the matter sector. 

The energy-momentum tensor is the one of a perturbed perfect fluid. The background contribution is given by a homogeneous perfect fluid with energy-momentum tensor
\be
\label{eq:Tmunu:background}
\bar{\cT}_{\mu\nu} = (\rho+P) u_{\mu} u_{\nu} + P g_{\mu\nu}\, ,
\ee
where
\begin{align}
\label{eq:umu:background}
 u^{\mu} \partial_{\mu} = a^{-1} \partial_\eta\, , \qquad u_{\mu} \rd x^{\mu} = - a \rd \eta
\end{align}
is the four velocity of a comoving fluid. The contribution from the perturbations reads
\begin{align}
\label{matp}
\delta \cT_{\mu\nu} =  (\delta \rho+\delta P) u_{\mu} u_{\nu} + \delta P g_{\mu\nu} +  (\rho+P) \left( \delta u_{\mu} u_{\nu} +  u_{\mu} \delta u_{\nu}\right) + P \delta g_{\mu\nu} + a^2 P \Pi_{\mu\nu}\, .
\end{align}
Here $\Pi_{\mu\nu}$ stands for the anisotropic pressure, which encodes the deviation from a perfect fluid. It satisfies $\Pi_{\mu\nu} u^{\nu} =0$ (hence $\Pi_{00}=0$ and $\Pi_{0i}=0$), and it is convenient to decompose it as
\be
\Pi_{ij} = \frac{1}{2} \Delta_{ij} \pi + {\DB}_{(i} \pi_{j)} + {\pi}_{ij}\, , \qquad \text{where} \qquad  \Delta_{ij} = {\DB}_{(i} {\DB}_{j)} - \frac{2}{3} {\gammaB}_{ij} \Delta
\label{eq:Deltaij:def}
\ee
with $\Delta={\DB}_i {\DB}^i$ and where the symmetrization is understood without factor $1/2$. Therefore, one can decompose the anisotropic stress tensor as
\be
{\DB}_i \pi^i =0\, , \qquad {\DB}_{i} \pi^i{}_j =0\, , \qquad \pi^i{}_i =0\, .
\ee
The perturbed four velocity vector is given by
\begin{align}
& \delta u^{\mu} \partial_{\mu}= -\frac{A}{a} \partial_\eta + \frac{\delta v^i}{a} \partial_i\, ,  \qquad \delta u_{\mu} \rd x^{\mu} = - a A \rd \eta + a (\delta v_i + N_i ) \rd x^i\, ,
\end{align}
where $\delta v_i = {\DB}_i  v  + v_i$. To summarize, the matter perturbations are encoded in the ten fields $(\delta \rho, \delta P, \delta v, v_i , \pi, \pi_i, \pi_{ij})$. Out of them, one can construct two useful gauge-invariant quantities that mix the gravitational and matter perturbations, namely the comoving curvature perturbation $\cR$ and the uniform curvature perturbation $\zeta$, respectively given by
\be
\label{eq:zeta:R}
\cR=-C-\mathcal{H}(v+B)\, ,\qquad \zeta=C-\mathcal{H}\frac{\delta\rho}{\etad{\rho}}\, .
\ee
Having described the geometry and matter sectors, we can now focus on their dynamics.

At the background level, let us introduce the equation-of-state parameter $\omega$ and the speed of sound $\cs$ defined as $P = \omega \rho$ and $ \cs^2 = \etad{P} / \etad{ \rho}$ respectively.
The background equations ${G}_{\mu\nu} = \kappa {T}_{\mu\nu}$ read
\begin{align}
\label{eq:Friedmann:1}
3\cH^2 & = \kappa a^2 \rho\, , \\
\cH^2 + 2 \etad{\cH}& = -  \kappa a^2 P\, ,\\
\etad{\rho} + 3\cH (1+\omega) \rho & =0\, ,
\label{eq:conservation}
\end{align}
where $\kappa=8\pi G/c^4$, while their linearized version, \ie~$\delta G_{\mu\nu} = \kappa \delta T_{\mu\nu}$, translate into the scalar constraint
\begin{align}
 3\cH \etad{C} -  \Delta \left[ C + \cH (B - \etad{E}) \right] & = \frac{\kappa}{2} a^2 \left(\delta \rho + 2 \rho A\right) , 
 \label{eq:1}
\end{align}
obtained from the $00$ component of Einstein's equations, and the momentum constraint 
\begin{align}
  {\DB}_i \left[ 2\cH A - 2 \etad{C}  - (2\etad{\cH} + \cH^2 ) B \right]  &  =  - \kappa a^2 \rho \left[ (1+\omega) {\DB}_i v + {\DB}_i B  \right]  , \\
 \frac{1}{2} \Delta ( E^{\prime}_i  - B_i ) - (2\etad{\cH} + \cH^2 ) B_i  & = - \kappa a^2 \rho \left[ (1+\omega)  v_i + B_i\right]  ,
\end{align}
that corresponds to the $0i$ component and that we have decomposed into a scalar and vector part. These are supplemented by the following dynamical equations for the scalar perturbations
\begin{align}
&  \etadd{C} +2 \cH \etad{C}  - (2\etad{\cH} + \cH^2) A- \cH \etad{A}  - \frac{1}{3} \Delta [A + C -(E' -B)' -2\cH (E' -B) ] = -  \frac{\kappa a^2}{2}\delta P   \, , \qquad \\
&  \Delta_{ij} \left(  \etadd{E} + 2 \cH \etad{E}  - \etad{B} - 2 \cH B - C - A \right) = \kappa a^2 P  \Delta_{ij} \pi \, ,
\end{align}
corresponding to the trace part and traceless scalar part of the $ij$ component, while its tensor and vectorial parts yield
\begin{align}
& \etadd{E_{ij}} + 2 \cH \etad{E_{ij}} - \Delta E_{ij} =  \kappa a^2 P  \pi_{ij}\, , \\
 \label{eq:2}
& \DB_{(i} \left\lbrace \left[E'_{j)} - B_{j)}\right]' + 2 \cH \left[E'_{j)} - B_{j)}\right] \right\rbrace  = \kappa P a^2 \DB_{(i} \pi_{j)} \, . 
\end{align}
Having summarized the kinematics and dynamics of the linear cosmological perturbation theory, we now focus on the Newtonian gauge.

\subsection{Scalar and tensor perturbations in the Newtonian gauge}

\label{A2}

In the rest of this work, for explicitness, we work in the Newtonian gauge, also called longitudinal gauge, where
\be
\label{NG}
E = B = 0\, , \qquad E_i =0\, .
\ee
Notice that the Newtonian gauge can also be realized with the conditions $E=B=0$ and $B_i=0$, used for instance in \cite{Peter:2013avv}. However, we choose here to impose $E_i=0$ to parallel the choice used in \cite{Pajer:2017hmb}.
 In this gauge, the gauge-invariant combinations~\eqref{GI1}-\eqref{GI2} reduce to 
 \be
\Psi = - C\, ,\qquad \Phi = A\, , \qquad \Phi_i = - B_i\, .
 \ee
In terms of these gauge-invariant variables, in the Newtonian gauge the metric~\eqref{pertmet} takes the form
\begin{align}
\label{pertmetnew}
\rd s^2 = a^2 \left\lbrace - (1+2 \Phi) \rd \eta^2  - 2 \Phi_i \rd \eta \rd x^i + \left[(1-2\Psi){\gammaB}_{ij} +  2 E_{ij} \right] \rd x^i \rd x^j\right\rbrace
\end{align}
where we recall that $E_i$ and $E_{ij}$ are subject to \Eq{eq:divergenceless:etc}.
Imposing the gauge fixing (\ref{NG}) in the perturbed Einstein equations, one obtains, in the scalar sector
\begin{align}
\label{eq:perturbedEinstein:scalar:Newton:in}
 \Delta \Psi  -3\cH (\etad{\Psi}+ \cH\Phi)  & = \frac{\kappa}{2} a^2 \delta \rho\, ,  \\
 {\DB}_i ( \etad{\Psi} + \cH \Phi  ) & = -  \frac{ \kappa}{2} a^2 \rho (1+\omega) {\DB}_i v \, ,\\
 \label{eq:perturbedEinstein:scalar:Newton:Psi''}
   \etadd{\Psi } + \cH (\Phi'+2 \etad{\Psi})  + (2\etad{\cH} + \cH^2) \Phi - \frac{1}{3} \Delta(\Psi - \Phi)  & = \frac{1}{2}  \kappa a^2 \delta P \, , \\
 \Delta_{ij} (\Psi - \Phi) & =  \kappa a^2 P \Delta_{ij}  \pi \, .
 \label{eq:perturbedEinstein:scalar:Newton:end}
\end{align}
The vector\footnote{Notice that \Eq{eom:Phi:pi} is different from equation (5.114) given in \cite{Peter:2013avv} since as mentioned above the Newtonian gauge has not been imposed in the same manner.} and tensor perturbations satisfy the following dynamical equations
\begin{align}
\label{eom:Phi:v}
D_{(i}\left[\Phi_{j)}'+2\mathcal{H}\Phi_{j)}\right]& =\kappa P a^2 D_{(i}\pi_{j)}\, , \\
\label{eom:Phi:pi}
\Delta\Phi_i + 4 \left(\cH'-\cH^2\right)\Phi_i & = -2\kappa a^2 \rho (1+\omega)v_i \, ,\\
\label{eom:C}
 \etadd{E_{ij}} + 2 \cH \etad{E_{ij}}  - \Delta E_{ij} & =  \kappa a^2 P \pi_{ij}\, .
\end{align}
Having presented the set of geometries we shall consider in this work, we now turn to the construction of soft perturbations by means of large gauge transformations.

\subsection{Generating soft perturbations}

\label{A3}

In this section, we revisit the construction of soft perturbations in the Newtonian gauge and show the enhanced infinite dimensional nature of the soft symmetry in this gauge.

\subsubsection{Metric perturbations}

Starting from the seed background FLRW metric
\be
\rd s^2 = \bar{g}_{\mu\nu} \rd x^{\mu} \rd x^{\nu} = a^2(\eta) \left( - \rd \eta^2 +  {\gammaB}_{ij} \rd x^i \rd x^j \right) ,
\ee
we consider a general diffeomorphism $x^\mu\to x^\mu-\xi^\mu$, which acts on the metric through the Lie derivative $\bar{g}_{\mu\nu}\to \bar{g}_{\mu\nu}+\cL_{\xi} \bar{g}_{\mu\nu} $, where
\be
\cL_{\xi} \bar{g}_{\mu\nu} = \nabla_{\mu} \xi_{\nu} + \nabla_{\nu} \xi_{\mu} = \xi^{\lambda} \partial_{\lambda} \bar{g}_{\mu\nu} + \bar{g}_{\mu\lambda} \partial_{\nu} \xi^{\lambda} + \bar{g}_{\nu\lambda} \partial_{\mu} \xi^{\lambda}\, .
\ee
At leading order, the metric thus transform as $\bar{g}_{\mu\nu}\to \bar{g}_{\mu\nu}+\cL_{\xi} \bar{g}_{\mu\nu} $ with
\begin{align}
\cL_{\xi} \bar{g}_{00} & =\xi^{\lambda} \partial_{\lambda} \bar{g}_{00} + \bar{g}_{0\lambda} \etad{ \xi}^{\lambda} + \bar{g}_{0\lambda} \etad{\xi}^{\lambda} = - 2 a^2 [(\etad{\xi^0)} + \cH \xi^0] \, ,\\
\cL_{\xi} \bar{g}_{0i} & = \xi^{\lambda} \partial_{\lambda} \cancel{\bar{g}_{0i} }+ \bar{g}_{0\lambda} \partial_i \xi^{\lambda} + \bar{g}_{i\lambda} \etad{\xi}^{\lambda} =  - a^2 [ {\DB}_i \xi^0 - {\gammaB}_{ij} \etad{(\xi^j)} ]\, , \\
\cL_{\xi} \bar{g}_{ij} & = \nabla_{i} \xi_{j} + \nabla_{j} \xi_{i} =  \left( {\DB}_i \xi_j + {\DB}_j \xi_i\right) + 2 a^2 {\gammaB}_{ij} \cH \xi^0\, ,
\end{align}
where $\bar{g}_{\mu\nu}$ denotes the background metric. The transformed metric thus reads
\beaq
\rd s^2 & = -a^2 \left[ 1 + 2 ( \etad{\xi}^0 + \cH \xi^0 )\right] \rd \eta^2 - 2a^2 ( {\DB}_i \xi^0 - {\gammaB}_{ij} (\xi^j)' ) \rd \eta \rd x^i  \\
&\;\;\;  + a^2  \left[ \left( 1 +  2  \cH  \xi^0 + \frac{2}{3} {\DB}_k \xi^k \right)  {\gammaB}_{ij}  +  \left({\gammaB}_{jk} {\DB}_i \xi^k + {\gammaB}_{ik} {\DB}_j \xi^k - \frac{2}{3}  {\gammaB}_{ij} {\DB}_k \xi^k\right) \right] \rd x^i \rd x^j
\eeaq 
where the space-space component has been split into trace and traceless contributions. Upon comparing with the perturbed metric (\ref{pertmet}), one can see that the gauge transformation has generated the metric perturbations
\begin{align}
A& =  \etad{\xi}^0 +\cH \xi^0\, , \\
 C& =\cH  \xi^0 + \frac{1}{3} {\DB}_k \xi^k\, ,\\
N_i  &= {\DB}_i B + B_i  =  -{\DB}_i \xi^0 + {\gammaB}_{ij} (\xi^j)'  \, , \\
h_{ij}-2C{\gammaB}_{ij}  & = 2 E_{ij} + 2 {\DB}_{(i} E_{j)} + 2 D_{(i} D_{j)} E  = {\gammaB}_{jk} {\DB}_i \xi^k + {\gammaB}_{ik} {\DB}_j \xi^k - \frac{2}{3}  {\gammaB}_{ij} {\DB}_k \xi^k\, .
\end{align}
Let us now restrict the analysis to the subset of diffeomorphisms that map the seed FLRW background into a perturbed FLRW universe in the Newtonian gauge (\ref{NG}). It implies that 
\begin{align}
 B_i  & =  -{\DB}_i \xi^0 + {\gammaB}_{ij} (\xi^j)' \, , \\
E_{ij} &=   \frac{1}{2}  \left( {\gammaB}_{jk} {\DB}_i \xi^k + {\gammaB}_{ik} {\DB}_j \xi^k - \frac{2}{3}  {\gammaB}_{ij} {\DB}_k \xi^k \right) .
\end{align}
The gauge fixing is supplemented by imposing the vanishing of the divergence of the vectorial and tensorial perturbations, which translate into
\begin{align}
\label{shift}
{\DB}_i B^i & = 0 \qquad \rightarrow  \qquad  {\DeltaB} \xi^0 = {\DB}_j (\xi^j)' \, ,  \\
\label{tens}
{\DB}_i E^{i}{}_{j} & = 0 \qquad \rightarrow \qquad {\DeltaB} \xi_i = - \frac{1}{3} {\DB}_i {\DB}_j \xi^j\, .
\end{align}
We can now write down the final metric we have constructed in terms of the gauge-invariant perturbations $(\Phi, \Psi, \Phi_i, E_{ij})$. It is given by (\ref{pertmetnew})
with 
\begin{align}
\label{pot1}
\Phi & =  \etad{(\xi^0)} +\cH \xi^0 \;, \\
\label{pot2}
 \Psi & = -\left(  \cH  \xi^0 + \frac{1}{3} D_k \xi^k\right)\, , \\
 \label{pot2p}
 \Phi_i  & = D_i \xi^0 - {\gammaB}_{ij} (\xi^j)' \, , \\
 \label{pot3}
E_{ij} &=   \frac{1}{2}  \left( {\gammaB}_{jk} {\DB}_i \xi^k + {\gammaB}_{ik} {\DB}_j \xi^k - \frac{2}{3}  {\gammaB}_{ij} {\DB}_k \xi^k \right)\, .
\end{align}
This is the geometry we shall study throughout this work. Note that, if $\xi^\mu$ is required to vanish at infinity, the only solution to \Eqs{shift} and~\eqref{tens} is given by $\xi^\mu=0$. This is because the gauge conditions entirely fix the system of coordinates in that case. General Relativity is invariant under space-time diffeomorphisms, which contribute boundary terms that vanish if $\xi^\mu$ vanishes at infinity. However, if $\xi^\mu$ does not vanish on the boundary, non-trivial solutions to \Eqs{shift} and~\eqref{tens} exist, and are referred to as ``soft'' diffeomorphisms. These equations have  to be understood as purely kinematical conditions on the family of soft diffeomorphisms. Before solving them, let us discuss the matter perturbations.

\subsubsection{Matter perturbations}

At the background level, the energy-momentum tensor is given by \Eq{eq:Tmunu:background}, which together with \Eq{eq:umu:background} leads to
\begin{align}
\bar{\cT}_{\mu\nu} \rd x^{\mu} \rd x^{\nu} = a^2 \left( \rho \; \rd \eta^2 + P {\gammaB}_{ij} \rd x^i \rd x^j\right)\, .
\end{align}
Being a rank-2 tensor, under the gauge transformation $x^\mu \to x^\mu-\xi^\mu$ it becomes $\cT_{\mu\nu}=\bar{\cT}_{\mu\nu}+\delta \cT_{\mu\nu}$ with
\be
\delta \cT_{\mu\nu} = \cL_{\xi} \bar{\cT}_{\mu\nu} = \xi^{\lambda} \partial_{\lambda} \bar{\cT}_{\mu\nu} + \bar{\cT}_{\mu\lambda} \partial_{\nu}\xi^{\lambda}  + \bar{\cT}_{\nu\lambda} \partial_{\mu}\xi^{\lambda} 
\ee
such that one obtains
\begin{align}
\label{eq:deltaT00:xi}
\delta \cT_{00} & = a^2 \xi^0 (  \etad{\rho} + 2 \cH \rho) + 2 a^2 \rho \etad{(\xi^{0})} = a^2 \left(  \xi^0 \etad{\rho} + 2 \rho \Phi \right) \\
\delta \cT_{0i} & = a^2 \left[ \rho \partial_i \xi^0 + P{\gammaB}_{ij} \etad{(\xi^j)} \right]  
= a^2 \rho \left[  (1+\omega) \gamma_{ij} (\xi^j)' + \Phi_i\right] \\
\delta \cT_{ij} & = a^2  \xi^0 (\etad{P} + 2 \cH P ){\gammaB}_{ij} + a^2 P ( {\gammaB}_{ik} \partial_j \xi^k +  {\gammaB}_{jk} \partial_i \xi^k + \xi^k \partial_k \gammaB_{ij})  = a^2  \left( \xi^0 \etad{P} - 2 P \Psi \right)  {\gammaB}_{ij} + 2a^2 PE_{ij}
\label{eq:deltaTij:xi}
\end{align}
where we have used \Eqs{pot1} and\eqref{pot2} in the second expression of $\delta \cT_{00}$, \Eq{shift} in the second expression of $\delta \cT_{0i}$, and \Eq{pot3} in the second expression of $\delta \cT_{ij}$. 

To identify the form of the matter perturbations, it is useful to write down the components of the perturbed energy-momentum tensor (\ref{matp}), which read
\begin{align}
\label{eq:deltaT00:Newton}
\delta \cT_{00} & =  a^2 (\delta \rho + 2 \rho \Phi ) \, ,\\
\delta \cT_{0i} & = - a^2 \rho  \left[ (1+\omega ) \left(  {\DB}_i v + v_i\right) -\Phi_i\right] \, , \\
\label{eq:deltaTij:Newton}
\delta \cT_{ij} & = a^2 (  \delta P - 2P \Psi) {\gammaB}_{ij} + a^2 P \left( 2 E_{ij} +  \Pi_{ij}\right) \, .
\end{align}
Upon identifying \Eqs{eq:deltaT00:xi}-\eqref{eq:deltaTij:xi} with \Eqs{eq:deltaT00:Newton}-\eqref{eq:deltaTij:Newton}, one finds for the energy and pressure perturbations
\be
\label{eq:matter:pert:xi}
\delta \rho = \xi^0 \etad{ {\rho}} \, ,\qquad \delta P = \xi^0 \etad{ {P}} \, .
\ee
Using \Eq{shift}, the vectorial sector is fixed by
\be
\label{eq:v:xi0}
D_i v + v_i  = - \gamma_{ij} (\xi^j)'  = - D_i \xi^0 + \Phi_i \qquad \rightarrow \qquad v = - \xi^0\, ,  \qquad v_i = \Phi_i\, . 
\ee
Finally, one immediately sees that the anisotropic pressure vanishes, \ie
\be
\Pi_{ij} = 0\, .
\ee
The first two identities are consistent with the fact that the energy density and the pressure are scalar quantities, hence they transform under Lie derivatives which reduce to standard derivatives in this case. Moreover, one notices that
\be
\frac{\delta P}{\delta \rho} = \frac{\etad{{P}}}{\etad{{\rho}}} = \cs^2\, ,
\ee
hence the soft perturbations are adiabatic. They also do not generate anisotropic pressure.
We have thus expressed all perturbations in terms of the components of the diffeomorphism $(\xi^0, \xi^i)$. In passing, we note that the curvature perturbations~\eqref{eq:zeta:R} is given by
\be
\label{eq:zeta:div:xi}
\zeta =-\mathcal{R}=\frac{1}{3}\DB_i \xi^i\, ,
\ee
where we have used \Eqs{pot2},~\eqref{eq:v:xi0} and~\eqref{eq:matter:pert:xi}. The curvature perturbations thus vanish when the spatial part of the soft modes is conserved ($\DB_i \xi^i=0$).

\subsection{Profiles of the soft diffeomorphism and soft modes}

\label{A4}

Our next task is to identify the soft generators that solve several conditions:
\begin{itemize}
\item {\textit{The diffeomorphism preserves the two gauge conditions~\eqref{shift} and~\eqref{tens}}}, which reads
\begin{align}
\label{shiftdiv}
  {\DeltaB} \xi^0 & = {\DB}_j (\xi^j )'\, , \\
  \label{tensdiv}
   {\DeltaB} \xi_i & = - \frac{1}{3} {\DB}_i {\DB}_j \xi^j\, .
\end{align}
\item {\textit{The perturbations generated by the diffeomorphism solve the linearized Einstein field equations.}} As shown in \App{FE}, upon imposing the background equations, \Eq{eq:perturbedEinstein:scalar:Newton:Psi''} is automatically satisfied while the remaining constraints can be summarized as follows. The scalar sector imposes that
\begin{align}
\label{EOM1}
\Delta D_i \xi^i & = 0\, , \\
\label{EOM2}
\Delta D_i \xi^0 & =0 \, ,\\
  \label{EOM4}
   \left({\DB}_{(i} {\DB}_{j)} - \frac{2}{3} {\gammaB}_{ij} {\DeltaB}\right) \left[ \etad{(\xi^0)}  + 2 \cH \xi^0 + \frac{1}{3} {\DB}_k \xi^k\right] & =0\, .
\end{align}
This last equation (\ref{EOM4}) will be central in our discussion on the residual low-multipole ambiguity. 
The tensor sector gives us
\begin{align}
\label{EOM5}
 \left({\gammaB}_{jk} {\DB}_i  + {\gammaB}_{ik} {\DB}_j  - \frac{2}{3}  {\gammaB}_{ij} {\DB}_k  \right) \left[\etadd{(\xi^k)} + 2\cH \etad{(\xi^k)}\right]  - \Delta \left({\gammaB}_{jk} {\DB}_i  + {\gammaB}_{ik} {\DB}_j  - \frac{2}{3}  {\gammaB}_{ij} {\DB}_k  \right)\xi^k = 0\, .
\end{align}
Finally, the vector sector imposes
\begin{align}
\label{EOM6}
 \Delta \left[ {\DB}_i \xi^0 - {\gammaB}_{ij} \etad{(\xi^j)}\right] & = 0\, ,\\
  \label{EOM7}
 \left\lbrace  a^2D_{(i}\left[{\DB}_{j)} \xi^0 - \gamma_{j)k}  (\xi^k)' \right] \right\rbrace '  & =0\, .
\end{align}
Upon solving this set of equations, one can identify the set of soft symmetries that are well-defined linearized diffeomorphisms, generating physical soft modes. 
\item {\textit{The two gauge conditions~\eqref{shift} and~\eqref{tens} are stable under the \textit{linearized} Lie bracket.}} This is automatically guaranteed when working at first order in the perturbations.
\end{itemize}

We now study two classes of solutions for the soft generators $(\xi^0, \xi^i)$ satisfying this set of conditions, namely the family of soft modes with vanishing divergence, \ie ${\DB}_i \xi^i =0$, and the family of soft modes with constant $\xi^i$. In what follows, \textit{we assume that the components of the diffeomorphism can be written with a separable ansatz} involving the product of a function of time and functions of spatial coordinates. Physically, this implies that none of the soft modes identified in what follows can be propagating. Note also that our goal is not to provide the most generic solutions to the above equations, but simply to display families of solutions that will be large enough to serve the purpose of the subsequent discussion.  

\subsubsection{Soft modes with vanishing divergence}
\label{sec:soft:mode:vanishing:divergence}

Consider first the family of soft symmetries whose spatial divergence is vanishing, \ie such that
\be
\label{eq:vanishing:divergence}
{\DB}_i \xi^i =0\, .
\ee
Plugging this relation into the gauge conditions, \Eqs{shiftdiv} and (\ref{tensdiv}) respectively impose that
\be
\Delta \xi_0 =0\, , \qquad \Delta \xi_i =0\, .
\ee
Using then the commutation between the spatial covariant derivatives, \Eqs{EOM1} and~\eqref{EOM2} are automatically satisfied, while \Eqs{EOM4} and (\ref{EOM5})  reduce to
\begin{align}
\label{EKRED1}
{\DB}_{(i} {\DB}_{j)}  \left[ \etad{(\xi^0)}  + 2 \cH \xi^0 \right] & =0\, , \\
\label{EKRED2}
 \left({\gammaB}_{jk} {\DB}_i  + {\gammaB}_{ik} {\DB}_j   \right) \left[\etadd{(\xi^k)} + 2\cH \etad{(\xi^k)}\right] &  = 0\, .
\end{align}
These equations can be solved with an ansatz of the form (note that other solutions may exist)
\begin{align}
\label{TDiff}
\xi^0 & = m(\eta) + g_i(\eta) x^i  + \frac{T(\vec{x})}{a^2} \\
\label{SDiff}
\xi^i & = \left[ \cC_1  +   \cC_2 \int^{\eta}\frac{\rd \tilde{\eta}}{a^2(\tilde{\eta})} \right] X^i (\vec{x})  + c^i(\eta) +  \Omega^{i}{}_j(\eta) x^j
\end{align}
where the functions $m(\eta)$, $g_i(\eta)$ and $c^i(\eta)$ are free, $(\cC_1, \cC_2)$ are constants, the function $T(\vec{x})$, $f(\vec{x})$ and the vector $X^i(\vec{x})$ satisfy
\begin{align}
\label{cond1}
\Delta  T & = 0 \, , \\
\label{cond2}
\Delta X_i & =0 \qquad \text{with} \qquad D_i X^i =0 \, ,
\end{align}
 while the time-dependent tensor $\Omega^i{}_j(\eta)$ is traceless and antisymmetric
\be
\Omega^i{}_i =0\, ,\qquad \Omega_{ij} = - \Omega_{ji}\, .
\ee
Notice that the time-dependence of $\xi^0$ and $\xi^i$ automatically solves \Eqs{EKRED1} and (\ref{EKRED2}). Moreover, the above conditions impose that $T$ and $X^i$ are respectively a harmonic function and a harmonic vector of the 3d euclidean space. Such functions and vectors can be split into growing and decaying branches of solution. Soft modes are only constructed thanks to the growing branch of the harmonic function and harmonic vectors. Indeed, when one computes the boundary charge associated to the diffeomorphism, the spatially decaying modes fall off and does not contribute to the charge, leading to a pure gauge transformation. However, the spatially-growing modes always contribute and therefore correspond to LGTs. We shall provide concrete examples of such diffeomorphisms in the following. 

At this stage, we still have to impose the remaining equations (\ref{EOM6}) and (\ref{EOM7}) coming from the vector sector.  A direct computation of the shift gives
\begin{align}
\label{vecprof}
\Phi_i = {\DB}_i \xi^0 - {\gammaB}_{ij} \etad{(\xi^j)} & =  g_i + \frac{1}{a^2} D_i T - \frac{\mathcal{C}_2}{a^2} X_i -\gamma_{ij} (c^i)' - \gamma_{ij}(\Omega^j{}_k)'x^k\, . 
\end{align}
Using \Eqs{cond1} and (\ref{cond2}), it is direct to check that \Eq{EOM6} is satisfied. Imposing the last equation (\ref{EOM7}), one finds that the term in the parenthesis is time-independent and the equation is automatically solved by the ansatz, leading to no further restriction. 

\subsubsection{Soft modes with constant spatial part}
\label{sec:soft:mode:constant:spatial:part}
Let us now consider the family of soft symmetries where
\be
\partial_{\eta} \xi^i =0 \, .
\ee 
In this case, the only equation coupling the components $\xi^i$ and $\xi^0$ is \Eq{EOM4}. Now, the gauge conditions reduce to
\begin{align}
\Delta \xi^0 =0 \qquad \text{and} \qquad   {\DeltaB} \xi_i & = - \frac{1}{3} {\DB}_i {\DB}_j \xi^j\, ,
\end{align}
which automatically solves \Eqs{EOM2} and (\ref{EOM6}). The remaining equations  (\ref{EOM1}) and (\ref{EOM4}) and the tensor equation (\ref{EOM5}) reduce to 
\begin{align}
\label{EOM11}
\Delta D_i \xi^i & = 0 \, ,\\
  \label{EOM44}
{\DB}_{(i} {\DB}_{j)}  \left[ \etad{(\xi^0)}  + 2 \cH \xi^0 + \frac{1}{3} {\DB}_k \xi^k\right] & =0\, , \\
\label{EOM55}
 \Delta \left({\gammaB}_{jk} {\DB}_i  + {\gammaB}_{ik} {\DB}_j    \right)\xi^k & = 0\, .
\end{align}
Since these equations involve third-order spatial derivatives acting on $\xi^i$, we can further assume for simplicity that the profile of $\xi^i$ is at most quadratic in $\vec{x}$. This restriction allows one to solve \Eqs{EOM11} and (\ref{EOM55}), while reducing (\ref{EOM44}) to
\be
{\DB}_{(i} {\DB}_{j)} \left[ \etad{(\xi^0)}  + 2 \cH \xi^0 \right]  =0\, .
\ee
This effectively decouples the $\xi^0$ and $\xi^i$ functions. Finally, the last equation (\ref{EOM7}) comes from the vector sector and reduces to 
\begin{align}
\label{EOM77}
\left(a^2 \DB_{(i} \DB_{j)} \xi^0 \right)'=0\, .
\end{align}
This is trivially satisfied with our ansatz. The equations for $\xi^0$ are the same as in the previous case so that the solution (\ref{TDiff})  identified above is valid. For the spatial component $\xi^i$, the restrictions we have imposed lead to the standard time-independent dilatation and special conformal transformations
\be
\xi^i = \lambda x^i + b^i x_j x^j - 2 b_j x^j x^i\, ,  
\ee
for which $\Delta D_i\xi^i =0$. Yet, other solutions can be constructed. In particular, one can check that the dilatation parameter and the special-conformal parameter can be upgraded to free time-dependent functions $\lambda(\eta)$ and $\beta(\eta)$ provided the time component $\xi^0$ is slightly modified.\footnote{For instance, one could generalize the dilatation to 
\be
\xi^0 = \lambda'(\eta) \gamma_{ij}x^i x^j\, , \qquad \xi^i = \lambda(\eta) x^i\, .
\ee} However, we will not use this extension here.

To summarize, the generators of the soft symmetries in the Newtonian gauge are given by 
\begin{align}
\label{TDifff}
\xi^0 & = m(\eta) + g_i(\eta) x^i  + \frac{T(\vec{x})}{a^2} \\
\label{SDifff}
\xi^i & = c^i(\eta)+ \Omega^i{}_{j}(\eta)x^j + \lambda x^i + b^i x_j x^j - 2 b_j x^j x^i  +  \left[ \cC_1  +   \cC_2 \int^{\eta}\frac{\rd \tilde{\eta}}{a^2(\tilde{\eta})} \right] X^i (\vec{x})  
\end{align}
where the functions $m(\eta)$, $c^i(\eta)$, $g_i(\eta)$ and $\Omega^i{}_j(\eta)$ are free, $(\lambda,b_i, \cC_1, \cC_2)$ are constants while the function $T(\vec{x})$  satisfies \Eq{cond1} and we have reabsorbed the freedom encoded in the function $f(\vec{x})$ in the time-dependent dipole.

\subsection{Examples of soft modes}

In this section, we provide explicit examples of soft diffeomorphisms that solve explicitly our different conditions. We first consider the case of the spatial component $\xi^i \partial_i$ and present examples for the two families of soft generators. Then, we discuss how our generalized soft symmetries allow one to construct soft cosmological modes that are new compared to the previous studies~\cite{Creminelli:2013mca, Pajer:2017hmb}.

\subsubsection{Anisotropies}

Let us first provide a few examples of the spatial soft generators with either vanishing divergence (they will be denoted as $X^i$) or constant spatial part (denoted as $Y^i$). Notice that as we have shown in the previous section, if one restricts to a time-independent spatial component for the soft diffeomorphism, one can choose $\xi^0 =0$ since the time and spatial components decouple. This is what we shall assume in the following. 

\begin{itemize}
\item The first family $X^i$ belongs to the class introduced in \Sec{sec:soft:mode:vanishing:divergence}: it is divergence free, \ie $D_i X^i =0$ and harmonic, \ie $\Delta X_i =0$,
hence it is associated to solenoidal solutions. As a concrete example of quadratic and cubic diffeomorphisms of this kind, one can consider
\begin{align}
X_{(1)}^{i} \partial_i & = x z \partial_x + y z \partial_y + \left[ \frac{(x^2+y^2)}{2} -z^2\right]  \partial_z\, ,  \\
X_{(2)}^{i} \partial_i & = x (y^2 -z^2)\partial_x - y (x^2 -z^2) \partial_y + z(x^2 - y^2) \partial_z\, .
\end{align}
By construction, this vector does not generate a curvature perturbation, see \Eq{eq:zeta:div:xi}. However, they do generate anisotropies. For instance, tensor perturbations of the form
\begin{align}
E^{(1)}_{ij} \rd x^i \rd x^j & =  (z\rd x + 2x \rd z) \rd x + (z \rd y + 2y \rd z)\rd y -2z \rd z^2\, , \\
\label{ANI}
E^{(2)}_{ij} \rd x^i \rd x^j & =  (y^2-z^2) \rd x^2 + (z^2-x^2)\rd y^2 + (x^2 - y^2)\rd z^2\, ,
\end{align}
see \Eq{pot3}.
They correspond to bona-fide large scalar anisotropies.
\item The second family $Y^i$ belongs to the solutions discussed in \Sec{sec:soft:mode:constant:spatial:part} and the example we consider is
\begin{align}
Y^{i} \partial_i & = x z \partial_x + y z \partial_y - \frac{1}{6} (x^2+y^2) \partial_z \, .
\end{align}
It is such that $D_i Y^i = 2 z$, hence from \Eq{eq:zeta:div:xi} it generates a non-constant dipolar curvature perturbation,
\be
\zeta =\frac{2}{3} z = \frac{2}{3}r \cos{\theta}\, ,
\ee
while the tensor perturbation reads
\be
E_{ij} \rd x^i \rd x^j  = \frac{1}{3}\left[ z \left( \rd x^2 + \rd y^2 -2 \rd z^2\right)  - 2 \left(x\rd x\rd z + y \rd y \rd z \right)  \right]\, .
\ee
\end{itemize}
These two profiles belong to the soft-mode families formally introduced in \cite{Pajer:2017hmb}, but they provide explicit examples of non-trivial anisotropies dressing the cosmological patch.

\subsubsection{New low-multipole modes with free time-dependence }
Let us now describe the new profiles of the perturbations that can be generated from our extended soft symmetries. Among the different possibilities, let us focus on the subsector generated by the soft generators
\begin{align}
\label{TDiffex}
\xi^0 & = m(\eta) + g_i(\eta) x^i\, ,  \\
\label{SDiffex}
\xi^i & = c^i(\eta) + \Omega^i{}_{j}(\eta) x^j\, .
\end{align}
It leads to the infinite family taking the form
\begin{align}
\label{SM0}
\Phi & = \left( m' + \cH m\right) + \left( g'_i + \cH g_i\right) x^i \\
 \Psi & = - \cH \left( m + g_i x^i \right) \\
  v & = - m - g_i x^i \\
  \delta \rho & = \rho' \left( m + g_i x^i  \right)   \\
    \delta P & = p' \left( m + g_i x^i \right) 
\end{align}
while the vector sector reads
\begin{align}
\label{SM1}
    v_i = \Phi_i & = g_i-\gamma_{ij}(c^j)' - \gamma_{ij}(\Omega^{j}{}_{k})' x^k
\end{align}
where we recall that $m(\eta)$, $g_i(\eta)$, $c_i(\eta)$ and $\Omega_{ij}(\eta)$ are free functions. All other gauge-invariant perturbations are vanishing. This family of modes is new and does not correspond to the first and second scalar Weinberg modes introduced in \cite{Weinberg:2003sw} and revisited in \cite{Pajer:2017hmb}. Let us now describe the effects of these soft modes on the cosmological patch. 

 \section{Physical effects on the geometry of the cosmological patch}

\label{D}

So far, we have provided a detailed description of the LGTs of the cosmological perturbations in the Newtonian gauge. The identification of the low-multipole soft modes with unconstrained time-dependence described above raises several questions. Do these new modes affect cosmological observables? Can they be considered as physical? How do they affect the geometry of the cosmological patch? 

Indeed, these soft modes have not been considered as physical in previous studies. In cosmology, the criteria used to decide the physicality of a soft mode constructed via LGTs were initially introduced by Weinberg in \cite{Weinberg:2003sw}. Working in Fourier space, a soft mode is said to be physical if it can be smoothly extended to finite momenta. In order for this to happen, a physical soft mode must satisfy the linearized equations of motion away from the zero momenta regime, i.e when $k\rightarrow0$. This is ensured by imposing more restrictive equations than the linearized Einstein equations. To be concrete, let us present one example. Consider \Eq{eq:perturbedEinstein:scalar:Newton:end} in the absence of anisotropic pressure, \ie $\pi=0$, which leads to 
\be
\label{crucial}
{\DeltaB}_{ij} (\Psi -\Phi) =   \left({\DB}_{(i} {\DB}_{j)} - \frac{2}{3} {\gammaB}_{ij} {\DeltaB}\right) \left[ \etad{(\xi^0)}  + 2 \cH \xi^0 + \frac{1}{3} {\DB}_k \xi^k\right] =0\, .
\ee
In Fourier space, this equation becomes
\be
\left( k_{(i} k_{j)} - \frac{2}{3} \gamma_{ij} k^2 \right) (\Psi-\Phi) = \left( k_{(i} k_{j)} - \frac{2}{3} \gamma_{ij} k^2 \right) \left[ \etad{(\xi^0)}  + 2 \cH \xi^0 + \frac{1}{3} {\DB}_k \xi^k\right] = 0\, .
\ee
The standard argument demands that one does not impose the field equation (\ref{crucial}), but instead the stronger version
\be
\Psi-\Phi =  \etad{(\xi^0)}  + 2 \cH \xi^0 + \frac{1}{3} {\DB}_k \xi^k = 0\, .
\ee
This ensures that even away from the limit $k\rightarrow 0$, the solution still holds. This procedure is problematic for several reasons. 

First, it automatically misses all the modes living in the kernel of the operator $\Delta_{ij}$. These are precisely (part of) the unconstrained soft modes discussed in this work. Second and more importantly, this criterion does not rely on a covariant, well-defined prescription, and is not directly related to the definition of large gauge transformations. In contrast, the geometrical and covariant criterion employed in other contexts (in particular gravitational waves and asymptotic symmetries \cite{Compere:2018aar, Harlow:2019yfa, Freidel:2020xyx, Ciambelli:2022vot}) is to recognize a long wavelength mode physical when it modifies the boundary structure of the spacetime, by turning on a non-vanishing boundary charge or by modifying a boundary observable. Therefore, in order to properly argue for the physicality of the new soft modes introduced here, it would be useful to clarify how they act on the boundary of the cosmological patch. 

To that end, a systematic approach would be to use the covariant phase-space methods to compute the Barnich-Brandt charges associated to our soft symmetries \cite{Compere:2018aar, Harlow:2019yfa, Freidel:2020xyx, Ciambelli:2022vot}. This task being cumbersome, it goes beyond the scope of this work and it will be presented in detail in a forthcoming publication \cite{NEXT}. See \cite{Kehagias:2016zry, Bonga:2020fhx, Enriquez-Rojo:2022onp} for efforts in this direction and \cite{DeLuca:2024asq, DeLuca:2024cjl} for works connecting the asymptotic and local point of view on these BMS and cosmological soft symmetries. Instead, the physical effects of the low-multipole soft modes on the boundary observables can be studied by computing the quasi-local energy and the angular momentum of the FLRW patch. 

\subsection{Quasi-local energy and angular momentum of the patch}

Since the equivalence principle allows one to locally erase the gravitational field, the notion of energy and angular momentum of a given gravitational system can only be quasi-local in General Relativity. The difficulty in identifying a well-defined notion of quasi-local energy (and angular momentum) has been a long-standing issue as evidenced by the many definitions available in the literature. See \cite{Szabados:2009eka} for a review. For isolated systems, such as stationary black holes, the Brown-York (BY) definition of the energy and angular momentum provides a satisfactory notion. However, for fully dynamical spacetimes such as the flat FLRW geometry, the BY definition fails to provide a sensible result. An extension of the BY definition was introduced by Epp in \cite{Epp:2000zr} and applied to the FLRW geometry by Afshar in \cite{Afshar:2009pi}, demonstrating its ability to account for the quasi-local energy of the FLRW patch. The fundamental concept on which this notion is built is the so-called mean-curvature frame associated to any closed $2$-surface $\cS$. In the following, we first review these objects and the definition of the mean curvature energy and angular momentum following \cite{Anco:2004bb}. Then we compute the quasi-local energy and angular momentum of the FLRW patch in the presence of soft monopoles and dipoles. 

\subsubsection{The mean-curvature frame and twist}

Consider a closed region $\cV$ in a four dimensional spacetime manifold $(\cM, g)$. Let us split the boundary $\partial \cV$ of that region as $\partial \cV = \Sigma_i \cup \cB \cup \Sigma_f$ where $(\Sigma_{i,f}, h)$ correspond to the two spacelike hypersurfaces at initial and final time with induced metric $h$, and $(\cB, \gamma)$ is a timelike hypersurface with induced metric $\gamma$. 

Now, consider the constant-time spacelike hypersurface $\Sigma_t$. Its cross-section with the timelike boundary $\cB$ defines a closed $2$-surface $\cS = \Sigma_t \cap \cB$. In the cosmological context, this closed $2$-surface represents the celestial sphere. Let us introduce the unit normal vector $n_{\mu} \rd x^{\mu}$ to $\Sigma_t$, \ie $n^{\mu} n_{\mu} = -1$, which is future pointing and the unit normal vector $s_{\mu} \rd x^{\mu}$ to $\cB$, \ie $s^{\mu} s_{\mu} = +1$, which is inward pointing and such that $g_{\mu\nu} n^{\mu} s^{\nu} =0$. The couple $(n,s)$ provides an orthonormal basis for the tangent space $T_p(\mathcal{S}^{\perp})$ and the metric on $\cS$ can be written as
\be
q_{\mu\nu} = g_{\mu\nu} + n_{\mu} n_{\nu} - s_{\mu} s_{\nu} \, ,
\ee
such that $q_{\mu\nu} n^{\mu} = q_{\mu\nu} s^{\mu} = 0$. Therefore, $q_{\mu}{}^{\nu}$ acts as a projector onto $\cS$ and one can define the covariant derivative on $\cS$ by $D_{\mu} = q_{\mu}{}^{\nu} \nabla_{\nu}$.

Now, since $\cS$ is a $2$-surface in a 4-dimensional spacetime $\cM$, its bending can be decomposed into its extrinsic curvature within the hypersurface $\Sigma_t$ or within the hypersurface $\cB$, respectively defined by
\begin{align}
K_{\mu\nu} (n)=  D_{\mu} n_{\nu}\, , \qquad K_{\mu\nu} (s)=  D_{\mu} s_{\nu}\, . 
\end{align}
In order to describe the bending of $\cS$ in $\cM$, one can thus define the so-called mean curvature vector $H$ and its dual $H_{\perp}$, given by
\begin{align}
\label{MCV}
H^{\mu} \partial_{\mu} & = \frac{1}{2} \left[ K(s) s^{\mu} \partial_{\mu} - K(n) n^{\mu}\partial_{\mu} \right] ,   \\
\label{MCVD}
H_{\perp}^{\mu} \partial_{\mu} & =\frac{1}{2} \left[  K(s) n^{\mu} \partial_{\mu} - K(n) s^{\mu}\partial_{\mu} \right] ,
\end{align}
where $K(n)$ and $K(s)$ are the traces of the extrinsic curvatures, \ie $K(n) = D_{\mu} n^{\mu}$ and $K(s) = D_{\mu} s^{\mu}$. Notice that, by construction, $g_{\mu\nu} H^{\mu} H^{\nu}_{\perp} =0 $. The couple $(H, H_{\perp})$ is known as the mean-curvature (orthogonal) frame of $\mathcal{S}^{\perp}$. Finally, the norm of both vectors is  given by
\begin{align}
g_{\mu\nu}H^{\mu} H_{\mu} = - g_{\mu\nu} H^{\mu}_{\perp} H^{\mu}_{\perp} = \frac{K^2(s) - K^2(n)}{4}\, ,
\end{align}
such that
\begin{align}
\label{norm}
 |H| = |H_{\perp}| =  \frac{1}{2}\sqrt{K^2(s) - K^2(n)}\, .
\end{align}
An interesting property of the couple $(H, H_{\perp})$ is that
\be
K(H_{\perp}) = D_{\mu} H^{\mu}_{\perp} =0 \, ,\qquad K(H) = D_{\mu} H^{\mu} = 2H^2 = 2H^2_{\perp}\, .
\ee 
Therefore, the dual mean-curvature vector $H_{\perp}$ singles out the normal direction w.r.t. $\cS$ along which the surface has a vanishing extrinsic curvature.
As we shall see, $H_{\perp}$ is the key object to construct an invariant notion of quasi-local energy for the closed $2$-surface $\cS$. Moreover, notice that $H_{\perp}$ is a straightforward generalization of the Kodama vector beyond spherical symmetry. In particular, it enjoys the similar property and can be used to track the presence of horizons within the geometry (see \cite{BenAchour:2025vur, BenAchour:2026zui} for an application to primordial axi-symmetric compact objects).   Another key object is the twist vector that encodes the angular momentum of the closed surface $\cS$ embedded in $\cM$. This mean-curvature twist (MCT) vector is defined as
\be
\label{eq:MCT:def}
\omega_{\mu} \rd x^{\mu} = H_{\alpha} D_{\mu} H^{\alpha}_{\perp} \rd x^{\mu}\, .
\ee (Notice that this definition differs from the one used in \cite{Anco:2004bb} by a factor $|H|^{-2}$).
If this vector does not vanish, the 2-surface $\cS$ carries an angular momentum. 
  
To fix ideas, let us compute the MCV and its dual for the flat FLRW geometry. Thus consider the metric
\begin{align}
\label{metback}
\rd s^2 = a^2(\eta) \left( - \rd \eta^2 + \rd r^2 + r^2 \rd \Omega^2 \right) 
\end{align}
where $\rd \Omega =  \rd \theta^2+\sin^2{\theta} \rd \varphi^2$ is the metric of the unit $2$-sphere.
Concretely, consider a bounded and expanding region in the FLRW geometry (\ref{metback}) and let us choose the boundary $\cS$ of constant $\eta$ and $r$ such that it admits the unit normals 
\be
\label{normalS}
n_{\mu} \rd x^{\mu} = a(\eta) \rd \eta\, , \qquad s_{\mu} \rd x^{\mu} = a(\eta) \rd r
\ee
and the induced metric on $\cS$ reads
\be
\label{indmetS}
q_{\mu\nu} \rd x^{\mu} \rd x^{\nu} = a^2(\eta) r^2 \left( \rd \theta^2 + \sin^2{\theta} \rd \varphi^2 \right).
\ee The MCV and its dual are given by
\begin{align}
H^{\mu} \partial_{\mu} = \frac{1}{a^2} \left( r^{-1} \partial_r - \cH \partial_{\eta} \right)\, , \qquad H_{\perp}^{\mu} \partial_{\mu} =- \frac{1}{a^2} \left( \cH \partial_r + r^{-1}\partial_{\eta} \right)\, .
\end{align}
As expected, the vanishing of its norm locates the apparent horizon of the FLRW cosmology. Concretely, one finds
\be
H_{\mu} H^{\mu} = - H^{\perp}_{\mu} H_{\perp}^{\mu} = \frac{1}{a^2} \left( \frac{1}{r^2} - \cH^2\right) ,
\ee
such that the MCV becomes null on the hypersurface $r = \pm \cH^{-1}$, recovering the well-known position of the cosmological horizon. Notice that a direct computation of the MCT shows that $\omega_{\mu} \rd x^{\mu} =0$, hence the cosmological patch does not carry any angular momentum, as expected. We can now introduce the notion of mean-curvature energy and angular momentum.

\subsubsection{Mean-curvature energy and angular momentum}

Let us first introduce the notion of mean-curvature energy $\epsilon = n_{\mu} H^{\mu}_{\perp}$ and mean-curvature momenta $p = s_{\mu} H^{\mu}_{\perp}$. Integrating them along the closed $2$-surface $\cS$, and setting the Newton constant to unity in this section for simplicity, one obtains
\begin{align}
E_{\text{MC}} (\cS)& =  \frac{1}{4\pi } \oint_{\cS} \rd^2 x \sqrt{q} \; \epsilon =- \frac{1}{8\pi } \oint_{\cS} \rd^2 x \sqrt{q} \; K(s)\, , \\
P_{\text{MC}} (\cS) & =  \frac{1}{4\pi } \oint_{\cS} \rd^2 x \sqrt{q} \; p  = -  \frac{1}{8\pi } \oint_{\cS} \rd^2 x \sqrt{q} \; K(n)\, ,
\end{align}
which reproduce respectively the well-known Brown-York quasi-local energy and the Brown-York momenta \cite{Brown:1992br}. 
The interesting outcome of these definitions is that the norm of the mean curvature vector combines both quantities such that
\be
|H_{\perp}| = \frac{1}{2}\sqrt{K^2(s) - K^2(n)}= \sqrt{\epsilon^2 - p^2}\, .
\ee
This elegant condensed form suggests interpreting this norm as encoding the rest quasi-local energy of the closed $2$-surface $\cS$. At this stage, it is natural to introduce the notion of mean-curvature quasi-local energy $\cE_{\text{MC}}$ by integrating the norm $|H_{\perp}|$ on the closed $2$-surface $\cS$ such that
\begin{align}
\label{EN}
\cE_{\text{MC}} (\cS) +  \cE^{\text{Ref}}_{\text{MC}} (\cS)= - \frac{1}{4\pi } \oint_{\cS} \rd^2 x \sqrt{q} |H_{\perp}| = - \frac{1}{8 \pi } \oint_{\cS} \rd^2 x \sqrt{q} \sqrt{K^2(s) - K^2(n)}
\end{align}
where we have used \Eq{norm} and we have introduced the standard reference term $\cE^{\mathrm{Ref}}_{\text{MC}} (\cS)$. This reference term corresponds to the mean-curvature energy computed in Minkowski space-times, within the adapted coordinate system, such that $\cE_{\text{MC}}$ is free from large-$r$ divergences. Such definition of quasi-local mass for a completely general spacetime geometry was introduced by Epp in \cite{Epp:2000zr} and studied in detail in \cite{Anco:2001gm, Anco:2001gk, Anco:2004bb} where it was dubbed the mean-curvature quasi-local mass. 

This definition of quasi-local energy $\cE_{\text{MC}}(\cS)$ was shown to provide consistent results in the context of flat and closed FLRW cosmological spacetimes, contrary to the standard Brown-York quasi-local energy alone \cite{Brown:1992br}. Indeed, as shown in \cite{Afshar:2009pi}, the BY energy vanishes for a flat FLRW patch, in conflict with the expectation. Concretely, consider a bounded and expanding region in the FLRW geometry (\ref{metback}) and let us choose the same boundary $\cS$ as introduced previously, i.e. such that it admits the unit normals \eqref{normalS} and the induced metric (\ref{indmetS}). Computing the Brown-York and the mean-curvature energy (and subtracting the appropriate reference term $\cE^{\text{Ref}}_{\text{MC}} (\cS) = - a(\eta) r$) for this FLRW patch, one obtains
\begin{align}
E_{\text{MC}} (\cS) & = E_{\text{BY}} = 0\, , \\
\label{en}
\cE_{\text{MC}} (\cS) &  = a(\eta) r \left[ 1- \sqrt{1-  r^2 \cH^2(\eta) }\right] .
\end{align} 
As emphasized in \cite{Afshar:2009pi}, the key reason for this discrepancy between the two definitions of energy lies in the fact that the BY one does not take into account all the components of the extrinsic curvature of the closed $2$-surface $\cS$. In particular, it only accounts for the component relative to the boundary $\cB$, thus ignoring the contribution $K(n)$.  While this is harmless for stationary spacetime for which $K(n) =0$, it provides an inconsistent result for fully time-dependent geometries for which $K(n) \neq 0$, as it is the case for any cosmological spacetimes. For this reason, the notion of mean-curvature quasi-local energy appears to be the most adapted concept of energy proposed so far to deal with sub-regions of cosmological spacetimes. We refer the reader to the work \cite{Oltean:2020mvt} for an application of this notion to cosmological perturbations.  

Notice that the quasi-local energy $\cE_{\text{MC}}(\cS)$, \ie \Eq{en},  is valid inside the cosmological horizon, \ie for $r \leqslant  \cH^{-1}$. As expected, the energy associated to a subregion of the flat FLRW spacetime is non zero, and vanishes only when the cosmic expansion/contraction is turned off, \ie when $\cH=0$, as shown initially in \cite{Afshar:2009pi}. Moreover, at the cosmological horizon, the quasi-local energy enclosed in the $2$-sphere representing the boundary of the observable universe is given by $\cE_{\text{MC}} (\cS)= \cH^{-1}$. Let us now turn to the mean-curvature angular momentum.

Consider the mean-curvature twist vector defined in \Eq{eq:MCT:def}. This vector is tangential to the closed surface $\cS$ and encodes the angular momentum of the system. Indeed, when the geometry admits a rotational Killing vector $\chi^{\mu} \partial_{\mu} = \partial_{\varphi}$, where $\varphi$ is the azimuthal coordinate, one can define the angular momentum as
 \be
 \label{MCVL}
 \cJ =\frac{1}{8\pi} \oint \rd^2 x \sqrt{q} \chi^{\mu} \omega_{\mu}\, .
 \ee
 It follows that if a given geometry possesses a non-vanishing MCT, it automatically admits a non-vanishing angular momentum and thus an axis of symmetry. Obviously, for the flat FLRW patch discussed above, the MCT twist vanishes such that the patch does not carry any angular momentum, \ie $\cJ_{\text{FLRW}}(\cS) =0$. 
  
Having discussed the definitions of mean curvature energy and angular momentum adapted to dynamical geometries, we can now study how the low-multipole soft modes modify the energy and angular momentum of the cosmological patch.

\subsection{Effect of the soft monopoles and dipoles}

In order to understand the effect of the low-multipole soft modes on the quasi-local properties of the cosmological patch, consider the soft generator $\xi^{\mu} \partial_{\mu}$ given in \Eqs{TDiffex}-\eqref{SDiffex}, \ie
\begin{align}
\label{TDiffex:2}
\xi^0 & = m(\eta) + g_i(\eta) x^i\, ,  \\
\label{SDiffex:2}
\xi^i & = c^i(\eta) + \Omega^{i}{}_{j}(\eta) x^j\, \qquad \text{with} \qquad \Omega_{(ij)} =0\, , \qquad \Omega^i{}_i =0\, , 
\end{align}
 which turns on the monopole and dipole contributions with free time-dependence. In order to study their effects on the quasi-local properties of the patch, let us further specialize the vector $g_i$ and the tensor $\Omega^i{}_{j}$ such that
 \begin{align}
 g_i =  \gamma_{ij} (c^j)' =\delta_i{}^z f(\eta)\, , \qquad   \Omega^x{}_{y} = - \Omega^y{}_{x}= J(\eta)\, .
 \end{align}
From \Eqs{pot1}-\eqref{pot3}, they generate the long-wavelength scalar and vector perturbations given in \Eqs{SM0}-\eqref{SM1}, \ie
\begin{align}
\Phi & = m'+ \cH m + \left(f'+ \cH f \right)z\, ,\\
\Psi & = - \cH m  - \cH f z \, ,\\
\Phi_x & = - J' y\, , \\
\Phi_y & =  J' x \, ,\\
\Phi_z & = 0\, ,
\end{align}
which contain only monopole and dipole terms. Notice that a direct computation shows that $E_{ij} =0$. The effect of these LWMs is best studied in spherical coordinates where they read
\begin{align}
\Phi (\eta, r, \theta)& = m'+ \cH m + \left(f'+ \cH f \right) r \cos{\theta}\, , \\
\label{psiprof}
\Psi(\eta, r, \theta) & = - \cH m  - \cH f r \cos{\theta}\, , \\
\label{spin}
\Phi_{\varphi}(\eta, r, \theta) & = -J' r^2 \sin^2{\theta}\, .
\end{align}
The resulting metric is given by 
\begin{align}
\label{pertmetnew:2}
\rd s^2 = a^2(\eta) \left\lbrace - \left[1+2 \Phi (\eta, r, \theta)\right] \rd \eta^2 + 2J' r^2 \sin^2{\theta} \rd \eta \rd \varphi + \left[1-2\Psi (\eta, r , \theta)\right] \left( \rd r^2 + r^2 \rd \Omega^2\right)\right\rbrace ,
\end{align}
see \Eq{pertmetnew}. Our goal now is to study the effects of these LWMs on the properties of the cosmological patch, \ie on the position of the horizon, its quasi-local energy and its angular momentum.
As discussed above, the key object to evaluate is the mean-curvature frame and its twist. 

Therefore, let us first describe the geometry of the cosmological patch. Consider a spacelike hypersurface $\Sigma_{\eta}$ with timelike unit normal 
\begin{align}
n_{\mu} \rd x^{\mu} & = a (1+  \Phi) \rd \eta\, ,  \qquad n^{\mu} \partial_{\mu} = - \frac{1}{a} \left[ (1-\Phi) \partial_{\eta} -J'\partial_{\varphi} \right]\, ,  
\end{align}
such that $ n_{\mu}n^{\mu} = -1$ and a timelike hypersurface $\cB_r$ with spacelike unit normal 
\begin{align}
s_{\mu} \rd x^{\mu} & = a(1-\Psi) \rd r\, , \quad \;\;\;\; s^{\mu} \partial_{\mu} = \frac{1+ \Psi}{a} \partial_r\, , 
\end{align}
such that $s_{\mu} s^{\mu} = +1$. This deforms \Eq{normalS}.
The boundary of the patch lies at the intersection $\cS = \Sigma_{\eta} \cap \cB_r$ and admits the induced metric $ q_{\mu\nu} = g_{\mu\nu} +n_{\mu} n_{\nu} - s_{\mu} s_{\nu}$ which takes the form
 \begin{align}
 q_{\mu\nu} \rd x^{\mu} \rd x^{\nu} = a^2 \left[ 2 J' r^2 \sin^2{\theta}  \rd \eta \rd \varphi  +  (1-2\Psi)r^2 \rd \Omega^2 \right] .
 \end{align}
 Computing the trace of the extrinsic curvature of $\cS$ w.r.t. $\Sigma_{\eta}$ and w.r.t. $\cB_{r}$, one finds at first order in the perturbations
 \begin{align}
 K(n) & = - \frac{2}{a} \left[ \cH - \Psi' - \cH (\Phi - \Psi) \right],\\
 K(s) & = \frac{2}{ar} \left[ 1+\Psi - r \partial_r \Psi \right].
 \end{align}
 Using the form of the unit normal vectors to the boundary $\cS$, one can build the MCV and its dual that read explicitly 
\begin{align}
H^{\mu} \partial_{\mu} & = -  \frac{\cH -\Psi' - \cH (2\Phi - \Psi)}{a^2} \partial_{\eta}+ \frac{1+ 2 \Psi - r \partial_r \Psi}{a^2 r}  \partial_r + \frac{\cH J'}{a^2} \partial_{\varphi}\, ,\\
H_{\perp}^{\mu} \partial_{\mu} & = -  \frac{1+  \Psi -\Phi - r \partial_r \Psi}{a^2 r}  \partial_{\eta}  + \frac{\cH - \Psi' - \cH (\Phi - 2\Psi)}{a^2} \partial_r +  \frac{J' }{a^2r} \partial_{\varphi}\, .
\end{align}
Equipped with this mean-curvature frame, we first compute the position of the cosmological horizon. The norm of the MCV is given by
\begin{align}
g_{\mu\nu} H^{\mu} H^{\nu} &  = \frac{1}{a^2 r^2 } \left\{ 1 + 2 ( \Psi -  r \partial_r \Psi ) - r^2 \left[ \cH^2 - 2 \cH \Psi' - 2 \cH^2(\Phi- \Psi)\right] \right\} .
\end{align}
By construction, the horizon corresponds to the locus of points at which the MCV becomes null, which is given by 
\be
r^2 = \frac{1 + 2 (\Psi -  r \partial_r \Psi)}{\cH^2 - 2 \cH (\Psi' + \cH \Psi)} \sim \frac{1}{\cH^2} \left[ 1 + \frac{2}{\cH } \Psi' - 2r \partial_r \Psi + 2 \Phi \right] \, .
\ee
Therefore, as expected, the soft monopoles $m(\eta)$ and the soft dipole $g_z(\eta) = f(\eta)$ entering in the expression of $\Psi$ modify the position of the cosmological horizon, demonstrating the first effect of this low-multipole modes on the cosmological patch. For $\Psi =0$, one recovers the standard result $r =  \cH^{-1}$. However, the corrected expression gives us a polynomial equation to solve that gives the explicit position of the horizon modified by the monopole $m$ and dipole $f$.

We can now compute the effect of the soft modes on the quasi-local energy of the patch. Using the definition (\ref{EN}) and subtracting the reference term w.r.t. the Minkowski background, one finds 
\begin{align}
\label{enFRPERT}
\cE^{\text{MC}} = a r \left( 1- \sqrt{|1-  r ^2 \cH^2|}\right) \left[ 1  + \frac{ r^2 \cH (\Psi' + \cH(\Phi+ \Psi)) - \Psi - r \partial_r \Psi}{1-r^2 \cH^2}  \right]  , 
\end{align}
which shows the correction to the quasi-local mass of the patch induced by the time-dependent monopole $m$ and dipole $g_i x^i$ contributions of $\Psi$. Recall that $\Psi$ being a perturbation, the last term of the last parenthesis has to remain small.  As expected, we recover the standard result derived by Afshar in \cite{Afshar:2009pi} for the flat FLRW geometry when setting $\Psi =0$. It is interesting to notice that the monopole $m$ and the dipole $g_i x^i$ shift the mass of the cosmological patch. This result is similar to the case of black hole perturbation theory, where the monopole and the polar-dipole perturbations of the Schwarzschild black-hole background induce a shift of the black-hole mass, \ie $M \rightarrow M + \delta M$. We also recover the fact that the axial dipole, $\Omega_{ij}x^j$, does not affect the mass but induces an angular momentum of the patch. This is analogous to the fact that the axial dipole perturbation transforms the Schwarzschild background to the slowly-rotating Kerr solution.\footnote{See \cite{BenAchour:2024skv} for an explicit computation of this zero-mode in black hole perturbation theory.} Since the time-dependence of the monopole and dipole contribution is free, it introduces an ambiguity in the shape and the energy of the FLRW patch. Indeed, one can always add or remove these monopole and dipole contributions, but at the expense of changing the energy enclosed in the patch. Notice that although the monopole contributions can be understood as a redefinition of the time and of the scale factor, this is no longer true for the dipole contribution.  In other words, fixing the geometry of the cosmological patch requires to fix this low-multipole ambiguity. 
 
 As a final point, let us now evaluate the twist vector in the presence of these soft modes. A lengthy computation shows that
 \begin{align}
 \omega_{\varphi} = \frac{J' r^2 \sin^2{\theta}}{a^2} \, .
 \end{align} 
 This automatically implies that the cosmological patch is now equipped with an angular momentum, the axis of rotation breaking the isotropy of the geometry. This effect also parallels what is known in black-hole perturbation theory. Indeed, perturbing the Schwarzschild black hole with an axial dipolar perturbation generates an angular momentum through a non-vanishing twist. The resulting gauge-inequivalent geometry is turned into the slowly-rotating Kerr black hole (similarly, the black-hole horizon is not modified at first order in the spin.) Therefore, the non-vanishing twist vector induced by the LGT (\ref{SDiffex}) is nothing else than the cosmological counterpart of this well-known effect in black-hole perturbation theory. 

At this level, it is clear that one cannot neglect the effect of these low multipole soft modes in the description of the cosmological patch as they affect the boundary observables, \ie  the quasi-local energy and/or the angular momentum of the cosmological patch we use to model the observable universe. Since the (linearized) Einstein's equations allow these low multipoles soft modes to have an arbitrary time-dependency, it follows that there is an inherent ambiguity in the description of the cosmological patch. In the next subsection, we comment on the interpretation of this ambiguity as a choice of cosmological frame. 

\subsection{Low-multipole freedom as a cosmological frame ambiguity}

In order to build a physical interpretation of the new freedom uncovered in the previous section in the low-multipole sector, let us contrast our result with the well-known case of asymptotically flat spacetimes (AFS). We refer the reader to the reviews \cite{Mitman:2024uss, Veneziano:2025ecv}. 
Radiative AFS are best described in the Bondi gauge. Denoting $(\theta, \varphi)$ the angular coordinate, $u$ the retarded time coordinate, $r$ the areal coordinate relative to the round $2$-sphere, the Bondi-Sachs metric reads
\begin{align}
\rd s^2 =  - V e^{2\beta} \rd u^2 - 2 e^{2\beta} \rd u \rd r + r^2 \gamma_{AB} \left( \rd x^A - U^A \rd u \right) \left( \rd x^B - U^B \rd u \right)
\end{align}
where the gauge $g_{rr} = g_{rA} =0$ has been imposed. In order for the metric to approach the Minkowski spacetime at large $r$, the functions $(V, \beta, U^A , \gamma_{AB})$ satisfy the fall-off conditions
 \be
V \rightarrow 1\;, \qquad U_A \rightarrow 0 \;,  \qquad \beta \rightarrow 0\;,  \qquad \gamma_{AB} \rightarrow \mat{cc}{1 & 0 \\ 0 & \sin^2{\theta}}\, .
\ee
Expanding these functions in powers of $1/r$, they can be decomposed into 
\begin{align}
V & = 1 -\frac{2M}{r} + \cO(r^{-2})\, , \qquad \beta  = \frac{\beta_0}{r} + \cO(r^{-2})\, , \\
U^A & = \frac{L^A}{r^2} + \cO(r^{-3})\, , \qquad \;\;\;\; \gamma_{AB} = q_{AB} + \frac{C_{AB}}{r} + \cO(r^{-2})\, , 
\end{align}
and the functions $(M, \beta_0, L^A, C_{AB})$ depend only on the coordinates $(u, \theta, \varphi)$. Imposing the Einstein equations and solving at each order in $1/r$, one obtains a solution space describing radiative AFS. Notice that a key difference with the treatment of cosmological perturbations is that while the dynamics is not linearized but exact, the resulting geometry is designed to hold far from the source, near null infinity, in the vicinity of $\cI^{+}$. As initially discovered by Bondi, Metzner and Sachs (BMS), this solution space enjoys a set of soft symmetries which extends the Poincar\'e isometries of Minkowski by a set of free angle-dependent time-translations and rotations respectively called supertranslations and superrotations \cite{Bondi:1960jsa, Bondi:1962rkt, Bondi:1962px}. The supertranslation corresponds to a time reparametrization that is angular dependent, \ie
 \be
 u \rightarrow u + \alpha (\theta, \varphi)
 \ee
where $ \alpha (\theta, \varphi)$ is a free function on the 2-sphere. This soft symmetry generates asymptotically non-trivial shear whose profile depends on the choice of the function $\alpha(\theta, \varphi)$. Therefore, it introduces an ambiguity in the description of AFS through the presence of unconstrained angular-dependent soft modes. These unconstrained modes give rise to the so-called memory observables, a major target of gravitational-wave astronomy for the decade to come. Indeed, analyzing the gravitational wave emitted by a binary black-hole merger event w.r.t. to a given AFS requires to fix a choice of the free function $\alpha (\theta, \varphi)$. Since it corresponds to a LGT, it is associated to a boundary charge whose value can be used to control this ambiguity and to characterize w.r.t. which AFS geometry one is considering this gravitational-wave event. This crucial point is known as the \textit{BMS frame ambiguity} and has been recognized as a key tool in the analysis of gravitational-wave templates~\cite{Mitman:2021xkq, Mitman:2022kwt, Iozzo:2021vnq, MaganaZertuche:2021syq, Compere:2019gft, Blanchet:2020ngx, Blanchet:2023pce}.  See \cite{Mitman:2024uss} for a pedagogical review on this point.
 
Similarly, we can understand the low-multipole soft modes uncovered in the previous section as introducing an inherent frame ambiguity in our description of the cosmological patch. Indeed, the time reparametrization
 \begin{align}
 & \eta \rightarrow \eta -  m(\eta) - g_i(\eta) x^i  \\
 & x^i \rightarrow x^i - c^i(\eta) - \Omega^{i}{}_{j}(\eta) x^j
 \end{align}
parallels the supertranslation and superrotation freedom of AFS discussed above, at least in the low-multipole sector. Analogously, a given model for the cosmological patch corresponds to a given choice of the functions $m(\eta)$, $g_i(\eta)$, $c^i(\eta)$ and $\Omega_{ij}(\eta)$. Therefore, analyzing cosmological observables requires to fix once and for all the geometry of the patch w.r.t. which this observable is interpreted. From this perspective, the analysis of various quantities that are usually decomposed in a multipolar expansion could be affected by this extra degree of freedom. Nevertheless, properly characterizing how this ambiguity could concretely manifest in cosmological observables requires further investigations that go beyond the scope of this work. 

At this point, it is worth emphasizing again that the approaches used for AFS and cosmological perturbation theory differ and a direct comparison would require a more involved treatment of the cosmological case. In particular, it would be desirable to reinvestigate the nature of the large gauge transformations within a foliation adapted to study radiative degrees of freedom. Such Bondi-like foliations have been worked out and are known as the observational gauge or the alternative light cone gauge \cite{Maartens:1995dx, Gasperini:2011us, Mitsou:2020czr}. A detailed investigation of the cosmological soft symmetries in these gauges will be presented in a forthcoming work. 

Before concluding this section, let us point out that the low-multipole soft modes could modify some standard statements in cosmological perturbation theory. Consider for instance a perturbed FLRW spacetime with vanishing anisotropic pressure, \ie $\pi=0$. In general, the vanishing of the scalar anisotropic pressure leads to $\Phi=\Psi$. However, the actual equation~(\ref{EOM4}) reduces to
\be
\label{crucial:2}
{\DeltaB}_{ij} (\Phi- \Psi) = 0\, ,
\ee
which involves a second-order traceless operator ${\DeltaB}_{ij}$, such that the equality of the two potentials $\Phi$
 and $\Psi$ is imposed only up to monopole and dipole contributions. Therefore, the resolution of this equation tells us that any solution to the linear Einstein equations can always be dressed by an arbitrary monopole and dipole because the field equations do not fix their time-dependence. This is why the equality $\Phi= \Psi$ commonly used in the literature should be taken with care, as it holds only at the quadrupolar order and beyond. As we have seen, the hidden soft modes allowed by this equation can affect the physical properties of the cosmological patch. This concludes the first part of this work focusing on the symmetries of the theory of cosmological perturbations. We now turn to the symmetries of the LWM cosmological solutions. 

\section{Killing symmetries for soft cosmological perturbations}

\label{B}

In this section, we switch the focus from the LGTs, which are symmetries of the theory, to the Killing symmetries of LWMs which are instead symmetries of a given solution of the linearized Einstein equations. 

The goal of this section is twofold. First, we wish to emphasize that all the LWMs generated by LGTs from the flat FLRW geometry naturally preserve the Killing symmetries of their FLRW seed. As we shall comment on in the following, this has interesting consequences for several approaches, among which the separate-universe one.  To that end, we derive the Lie-dragged Killing symmetries for the soft cosmological perturbations presented in \Sec{A}. Second, we highlight that these symmetries can be used to solve technical problems, such as characterizing the effects of the these large-scale cosmological perturbations on test probes such as particles or fields.

As a starting point, we review the (conformal) isometries and the hidden symmetries of the homogeneous and isotropic FLRW geometry before Lie-dragging them to the soft sector. 

\subsection{Lie dragging Killing symmetries}

\label{B1}

Consider a seed manifold and its metric $(\bar{M}, \bar{g})$ and  let us construct a new gauge-inequivalent solution $(M, g)$ by a soft diffeomorphism $\chi^{\mu} \partial_{\mu}$ such that the new metric is given by
\be
g = \cL_{\chi} \bar{g}\, .
\ee
We consider the expansion of the generator such that one can write $\mathcal{L}_\chi = \mathrm{Id}+\epsilon \mathcal{L}_\xi$.
 Then, the new metric decomposes into 
\be
g = \bar{g} +\epsilon  \cL_{\xi} \bar{g}\, .
\ee 
In order to study how the Killing symmetries are transferred from the seed to the new solution, we first need to transform the Levi-Civita connection under the Lie derivative. Under a diffeomorphism generated by the vector field $\xi^{\mu} \partial_{\mu}$, in \App{app:commutator:Lie:covariant:derivative} we show that the Christoffel symbols transform as $\Gamma^{\lambda}{}_{\mu\nu} \to \bar{\Gamma}^{\lambda}{}_{\mu\nu} +\delta_{\xi} \bar{\Gamma}^{\lambda}{}_{\mu\nu} $ where
\begin{align}
\label{formula}
\delta_{\xi} \bar{\Gamma}^{\lambda}{}_{\mu\nu} = \bar{\nabla}_{\mu} \bar{\nabla}_{\nu} \xi^{\lambda} + \bar{R}^{\lambda}{}_{\nu\rho\mu} \xi^{\rho}\, .
\end{align}
With this transformation at hand, one can now verify that given a generator of a symmetry on the seed, the Lie-dragged version of this generator is a symmetry of the new geometry. To start with, let us assume that the background metric $\bar{g}$ admits a conformal Killing vector $\bar{X}^{\mu} \partial_{\mu}$, which is defined by
\begin{align}
& \bar{\nabla}_{\mu} \bar{X}_{\nu} + \bar{\nabla}_{\nu} \bar{X}_{\mu} = \bar{\Omega} \bar{g}_{\mu\nu} \, .
\end{align}
Consider now the Lie-dragged Killing vector $X_{\mu} = \cL_{\chi}\bar{X}_{\mu}$. Using \Eq{formula}, it is straightforward to show that 
\begin{align}
\label{eq:Lie:Dragged:KV}
& \nabla_{\mu} X_{\nu} + \nabla_{\nu} X_{\mu} = \cL_{\chi} \left( \bar{\nabla}_{\mu} \bar{X}_{\nu} + \bar{\nabla}_{\nu} \bar{X}_{\mu} \right) = \Omega g_{\mu\nu} 
\end{align}
where $\Omega = \bar{\Omega}+ \cL_{\xi} \bar{\Omega}$. This is true for any Killing symmetry, and at all orders in $\epsilon$.

 Let us assume that the background metric $\bar{g}$ also admits a Killing-Yano (KY) two-form $\bar{Y}_{\nu\alpha}$ and a rank-$2$ Killing tensor (KT) $\bar{K}_{\mu\nu}$ which are respectively defined by
\begin{align}
\label{eq:KY:def}
& \bar{\nabla}_{\mu} \bar{Y}_{\nu\alpha} + \bar{\nabla}_{\nu} \bar{Y}_{\mu\alpha} =0\, , \\
\label{eq:KT:def}
& \bar{\nabla}_{\mu} \bar{K}_{\nu\alpha} + \bar{\nabla}_{\nu} \bar{K}_{\mu\alpha} + \bar{\nabla}_{\alpha} \bar{K}_{\mu\nu} =0\, .
\end{align}
Consider the Lie-dragged versions of these background symmetry generators given by 
\be 
Y_{\mu\nu} = \cL_{\chi} \bar{Y}_{\mu\nu}\, , \qquad K_{\mu\nu} = \cL_{\xi} \bar{K}_{\mu\nu}\, . 
\ee
Using \Eq{formula}, it is direct to show that 
\begin{align}
\label{KYCOM}
& \nabla_{\mu} Y_{\nu\alpha} + \nabla_{\nu} Y_{\mu\alpha} = \cL_{\chi} \left( \bar{\nabla}_{\mu} \bar{Y}_{\nu\alpha} + \bar{\nabla}_{\nu} \bar{Y}_{\mu\alpha} \right) = 0 \, ,\\
& \nabla_{\mu} K_{\nu\alpha} + \nabla_{\nu} K_{\mu\alpha} + \nabla_{\alpha} K_{\mu\nu} = \cL_{\chi} \left( \bar{\nabla}_{\mu} \bar{K}_{\nu\alpha} + \bar{\nabla}_{\nu} \bar{K}_{\mu\alpha} + \bar{\nabla}_{\alpha} \bar{K}_{\mu\nu}  \right) = 0\, .
\end{align}
For completeness, we present the explicit derivation of \Eq{KYCOM} in \App{KYapp}.
Therefore, as expected, all the explicit and hidden Killing symmetries of the background are preserved under a large gauge diffeomorphism generating the new gauge-inequivalent solution. 

Let us make a few remarks here. First, notice that for conformal Killing vectors, we can alternatively use a formulation in terms of the Lie derivative that avoids the use of \Eq{formula}. This alternative computation is presented in \App{CKVapp}. 
Second, the algebra of (conformal) Killing vectors is preserved under the action of the soft diffeomorphism. To see this, let us consider two given vector fields of the background, \ie $\bar{X}^{\mu} \partial_{\mu}$ and $\bar{Y}^{\mu} \partial_{\mu}$, which satisfy
\be
\label{eq:com:prod}
[\bar{X}, \bar{Y}] = \left( \bar{X}^{\mu} \partial_{\mu} \bar{Y}^{\alpha} - \bar{Y}^{\mu} \partial_{\mu} \bar{X}^{\alpha}\right) \partial_{\alpha}  =  \bar{Z}\, .
\ee
Consider the two Lie-dragged vectors $X = \cL_{\chi} \bar{X} = \bar{X}+\epsilon [\xi,\bar{X}]$ and $Y = \cL_{\chi} \bar{Y}+\epsilon [\xi,\bar{Y}]$. Their algebra reads\footnote{The form~\eqref{eq:com:prod} is the commutator for the product $(A\cdot B)^\mu = A^\nu \partial_\nu B^\mu$. Note that this product is not associative, \ie $(A\cdot B)\cdot C \neq A\cdot(B\cdot C\cdot)$. This is why, when going from \Eq{eq:com:1} to \Eq{eq:com:2}, one should not omit the terms of the form $(A\cdot B)\cdot C - A\cdot(B\cdot C\cdot)$, whose contribution nonetheless cancels out.}
\begin{align}
\label{eq:com:1}
[X, Y] & = [\bar{X}, \bar{Y}] + \epsilon [ [\xi,\bar{X}], \bar{Y}] + \epsilon [\bar{X} ,  [\xi,\bar{Y}]] + \cO(\epsilon^2) \\
\label{eq:com:2}
& = \bar{Z} + \epsilon [\xi, [\bar{X}, \bar{Y}]] + \cO(\epsilon^2) \\
& = \bar{Z} + \epsilon \cL_{\xi} \bar{Z} + \cO(\epsilon^2) \\
& = Z + \cO(\epsilon^2) 
\end{align}
where $Z = \cL_{\chi} \bar{Z}=  \bar{Z} + \epsilon \cL_{\xi} \bar{Z}$. Therefore, up to first order in the perturbation, the soft diffeomorphism preserves the conformal Killing algebra inherited from the background metric we started with. It follows that even if the Killing generators are deformed by the presence of the soft modes, their algebraic relations are preserved. We now turn our attention to the geodesic flow.

\subsection{Killing symmetries of the FLRW geometry}

\label{B2}

Consider the FLRW geometry that corresponds to our background metric 
\be
\rd s^2 = a^2(\eta) \left( - \rd \eta^2 + {\gammaB}_{ij} \rd x^i \rd x^j \right)  \, .
\ee
This geometry is conformally flat such that one can write $\rd s^2 = a^2(\eta) \tilde{\eta}_{\mu\nu} \rd x^{\mu} \rd x^{\nu}$ where $\tilde{\eta}_{\mu\nu}$ is the Minkowski metric. In particular, this implies that the Weyl tensor vanishes and that the background is algebraically special of Petrov type O. Let us first analyze the conformal Killing isometries of the FLRW geometry.

\subsubsection{Conformal Killing isometries}

As shown in \cite{Alvarez:2020pxc}, given a Conformal Killing Vector (CKV) $\tilde{\chi}^{\mu}\partial_{\mu}$ of the Minkowski geometry, one can construct a CKV $\chi^{\mu} \partial_{\mu}$ of the FLRW geometry through the relation\footnote{For two metrics related through $g_{\mu\nu} = \Omega \tilde{g}_{\mu\nu}$, the relation between CKVs generalizes to~\cite{Alvarez:2020pxc}
\begin{align}
\chi^{\mu} \partial_{\mu} = \tilde{\chi}^{\mu} \partial_{\mu}\, , \qquad \chi_{\mu} \rd x^{\mu} = \Omega^2 \tilde{\chi}_{\mu} \rd x^{\mu} \, .
\end{align}}
\begin{align}
\chi^{\mu} \partial_{\mu} = \tilde{\chi}^{\mu} \partial_{\mu} \, ,\qquad \chi_{\mu} \rd x^{\mu} = a^2(\eta) \tilde{\chi}_{\mu} \rd x^{\mu} \, .
\end{align}
It follows that the FLRW geometry admits the same conformal isometry group as Minkowski, namely $\SO(4,2)$, generated by the fifteen CKVs. These CKVs can be compactly written as
\begin{align}
\label{CKVFLRW}
\chi^{\mu} \partial_{\mu} = \left[  T^{\mu} + \Omega^{\mu}{}_{\nu} x^{\nu} + \omega x^{\mu}  + 2 \beta_{\lambda} x^{\lambda}  x^{\mu} - \beta^{\mu} (\eta_{\alpha\beta} x^{\alpha} x^{\beta}) \right] \partial_{\mu}
\end{align}
where $T^{\mu}$, $\Omega^{\mu}{}_{\nu}$, $\omega$ and $\beta^{\mu}$ are constant parameters encoding respectively the translations, rotations and boosts, dilatations and special-conformal transformations. Notice that $\Omega^{\mu}{}_{\nu}$ is an anti-symmetric and traceless tensor. 
A crucial point is that while these vectors are indeed CKVs of the flat FLRW geometry, the splitting between Killing Vectors (KVs), \ie CKVs satisfying $\nabla_{\mu} \chi^{\mu} =0$, and CKVs for which $ \nabla_{\mu} \chi^{\mu} \neq 0$, is different from their Minkowski counterpart. Indeed, while the translations, the rotations and the boosts are KVs of the Minkowski metric, this is no longer true in the FLRW geometry since in these cases
\begin{align}
\tilde{\nabla}_{\mu} \tilde{\chi}^{\mu} & = \partial_{\mu} \tilde{\chi}^{\mu} =0 \\
\text{but}\quad\nabla_{\mu} \chi^{\mu} & = \partial_{\mu} \chi^{\mu} + \Gamma_{00}^0\chi^0 + \Gamma_{i0}^i \chi^0 = 4 \cH \chi^0\, ,
\end{align}
where we have used that the matrices $\Omega^{\mu}{}_{\nu}$ are traceless.
Only the spatial translations and the rotations remain Killing vectors for the FLRW geometry, but the time translations and the boosts for which $\chi^0 \neq 0$ become conformal Killing vectors\footnote{Let us check explicitly that the time translation $\partial_{\eta}$ is indeed a CKV of the FLRW geometry. It acts on the metric as follows
\be
\cL_{\partial_{\eta}} g_{\mu\nu} = g'_{\mu\nu} = 2 \cH g_{\mu\nu}\, ,
\ee
which shows that the non-stationarity of the FLRW geometry does not spoil the time translation as a symmetry, but it modifies it such that it acquires a non-vanishing divergence, making it a CKV instead of a KV. }. Therefore, while the conformal isometry group remains the same, the isometry group generated by Killing vectors is smaller, reducing from ten to six the number of generators.

Let us now give a concrete example to make the above discussion more explicit. Consider the FLRW background geometry with a scale factor evolving as a power law
\be
\label{pow}
\rd s^2 = a^2(\eta) \left( - \rd \eta^2 + {\gammaB}_{ij} \rd x^i \rd x^j \right)\, , \qquad\text{where}\ \ a(\eta) = \eta^p\, .
\ee 
This geometry admits six KVs generating spatial translations and rotations, and given by
\begin{align}
T^i \partial_i = & \{ \partial_x, \partial_y, \partial_z\}\, , \\
R^i \partial_i = & \{ y \partial_z - z \partial_y, x \partial_z - z \partial_x,  x\partial_y - y \partial_x \} \, ,
\end{align}
but it also admits a set of nine CKVs. First, one has four special-conformal transformations and one dilatation generators given by
\begin{align}
S_1^{\mu} \partial_{\mu} & = \frac{1}{2} (\eta^2 - x^2 - y^2 -z^2 ) \partial_x +   \eta x \partial_{\eta}  + x \left( y \partial_y + z \partial_z \right)\, , \\
S_2^{\mu} \partial_{\mu} & =  \frac{1}{2} (\eta^2 - x^2 - y^2 -z^2 ) \partial_y +  \eta y \partial_{\eta}   + y \left( x \partial_x + z \partial_z \right)\, , \\
S_3^{\mu} \partial_{\mu} & =  \frac{1}{2} (\eta^2 - x^2 - y^2 -z^2 ) \partial_z +  \eta z \partial_{\eta}   + z \left( x \partial_x + y \partial_y \right)\, , \\
S_4^{\mu} \partial_{\mu} & = \frac{1}{2} (\eta^2 - x^2 - y^2 -z^2 ) \partial_{\eta}  + \eta \left( x \partial_x + y \partial_y + z \partial_z \right)\, ,  \\
{\DB}^{\mu} \partial_{\mu} & = \eta \partial_{\eta}  + x \partial_x  + y \partial_y + z \partial_z  \, .
\end{align}
Then, one has three boosts and one (would-be) time translation given by
\begin{align}
E^{\mu} \partial_{\mu} & =  \partial_{\eta} \, ,   \\
B_1^{\mu} \partial_{\mu} & = \eta \partial_x + x \partial_{\eta}\, ,  \\
B_2^{\mu} \partial_{\mu} & = \eta \partial_y + y \partial_{\eta} \, , \\
B_3^{\mu} \partial_{\mu} & = \eta \partial_z + z \partial_{\eta}  \, .
\end{align}
It is direct to check that all these vector fields generate conformal rescaling of the FLRW metric (\ref{pow}).
By construction, all the fifteen CKVs lead to deformed CKVs in the perturbed FLRW geometry. But before presenting their explicit form, let us expand on the other symmetries admitted by the FLRW background geometry.

\subsubsection{Killing-Yano and Killing tensors}

Higher symmetries generated by tensors are usually dubbed hidden symmetries as they do not act directly on the metric as (conformal) Killing vectors by the Lie derivative. However, they do generate explicit conserved charges for the geodesic motion on the associated geometry.

A first kind of hidden symmetry is the one generated by rank-$2$ and rank-$3$ Killing-Yano tensors. We shall be only interested in rank-$2$ KY tensors in what follows. These are anti-symmetric tensors satisfying \Eq{eq:KY:def}. The conditions for the existence of such objects have been studied in detail in the literature, see \cite{Stephani:2003tm}. Spacetimes with constant curvature admit ten such KY tensors. However, for non-Einstein spaces, the number of KY tensors is reduced. In the case of the FLRW geometry, one can identify four rank-$2$ KY tensors of the form~\cite{Popa:2004mq}
\begin{align}
\bar{Y}^{(1)}_{ij} \rd x^i \wedge \rd x^j & = a^3 \rd x \wedge \rd z  \, ,\\
\bar{Y}^{(2)}_{ij} \rd x^i \wedge \rd x^j & = a^3 \rd y \wedge \rd z  \, ,\\ 
\bar{Y}^{(3)}_{ij} \rd x^i \wedge \rd x^j & =a^3 \rd x  \wedge \rd y  \, ,\\
\bar{Y}^{(4)}_{ij} \rd x^i \wedge \rd x^j & =  a^3 \left( x \rd y \wedge \rd z + y \rd z \wedge \rd x + z  \rd x \wedge \rd y\right)  \, ,
\end{align}
which can be written in a compact form as 
  \be
  \label{KKY}
 \bar{Y}^{(4)}_{ij} = a^3 \epsilon_{ijk} x^k \qquad\text{and}\qquad  \bar{Y}^{(I)}_{ij} = a^3 \epsilon_{ijk} f_{(I)}^k \qquad  \text{with} \qquad f^k_{(I)} = \left\{
    \begin{array}{ll}
        f^k_1 = \delta^k_x &  \\
        f^k_2 = \delta^k_y & \\
        f^k_3 = \delta^k_z
    \end{array}
\right.
  \ee
for $I=1,\,2\,,3$.
  By construction, these four KY tensors induce four rank-$2$ KY tensors for the perturbed FLRW geometry.
  
Now, the second type of hidden symmetry is generated by Killing tensors. We shall again restrict our attention to rank-$2$ KTs, which are defined by \Eq{eq:KT:def}. A spacetime geometry admits at most fifty linearly independent KTs, this maximal number being achieved only by spacetimes with constant curvature. The existence of a KY tensor $Y_{\mu\nu}$ naturally induces the existence of a Killing tensor of the form
  \be
  K_{\mu\nu} = Y_{\mu\alpha} Y^{\alpha}{}_{\nu}\, .
  \ee
Nevertheless, KTs are not all descending from KY tensors.

In the FLRW case, one can exhibit twenty two KTs among which four descend from the four KY tensors presented above. In the present case, the KTs associated to the KY tensors (\ref{KKY}) are given by
\begin{align}
\bar{K}^{(I)}_{ij} = \bar{g}^{\alpha\beta} \bar{Y}^{(I)}_{i\alpha} \bar{Y}^{(I)}_{\beta j} =  \bar{g}^{k\ell} \bar{Y}^{(I)}_{ik} \bar{Y}^{(I)}_{\ell j}  = - a^4 {\gammaB}^{k\ell} \epsilon_{kim} \epsilon_{\ell j n} f^m_{(I)} f^n_{(I)}\, .
\end{align}
  One can further write them as
    \begin{align}
  \bar{K}^{(1)}_{ij} = -a^4 \mat{ccc}{0 & 0& 0 \\
  0& 1 & 0 \\
  0& 0& 1}\, , \qquad   \bar{K}^{(2)}_{ij} = -a^4 \mat{ccc}{1 & 0& 0 \\
  0& 0 & 0 \\
  0& 0& 1} \, ,\qquad   \bar{K}^{(3)}_{ij} = -a^4 \mat{ccc}{1 & 0& 0 \\
  0& 1 & 0 \\
  0& 0& 0}\, .
  \end{align}
  Nevertheless, the twenty two KTs of the FLRW geometry are trivial, in the sense that they correspond to the metric and to the twenty one symmetrized products between the six linearly independent Killing vectors $\chi^{\mu} \partial_{\mu}$ of the geometry. Concretely, they can all be written as
  \begin{align}
  K_{\mu\nu} = A g_{\mu\nu} + B \chi_{(\mu} \chi_{\nu)}
  \end{align}
  where $(A,B)$ are constants. It follows that they do not generate any additional conserved charges compared to the one already obtained from the isometries. For this reason, we shall not focus on these Killing tensors in the following. Nevertheless, we shall see that the absence of non-trivial KTs does not prevent us from algebraically integrating the geodesic motion. We can now derive the Killing symmetries in the presence of the adiabatic cosmological perturbations.

  \subsection{Killing symmetries for the soft cosmological perturbations}
  
  \label{B3}
  
 Consider the perturbed FLRW geometry, given by \Eq{pertmetnew:2} in the Newtonian gauge, and that we recall here for convenience,
 \begin{align}
\rd s^2 = a^2 \left\lbrace - (1+2 \Phi) \rd \eta^2  + 2 \Phi_i \rd \eta \rd x^i + \left[(1-2\Psi){\gammaB}_{ij} +  2E_{ij} \right] \rd x^i \rd x^j\right\rbrace\, .
\end{align}
We also recall that, in terms of the soft diffeomorphism, the perturbations admit the profiles~\eqref{pot1}-\eqref{pot3}, namely
\begin{align}
\label{pot}
\Phi & =  \etad{(\xi^0)} +\cH \xi^0 \, ,\\
 \Psi & = -\left(  \cH  \xi^0 + \frac{1}{3} {\DB}_k \xi^k\right)\, , \\
 \Phi_i & = D_i \xi^0 - {\gammaB}_{ij} (\xi^j)'\, ,  \\
E_{ij} & =  \frac{1}{2} \left( {\gammaB}_{jk} {\DB}_i \xi^k + {\gammaB}_{ik}{\DB}_j \xi^k - \frac{2}{3}  {\gammaB}_{ij} {\DB}_k \xi^k \right)\, .
\end{align}
Let us construct the different generators of the explicit and hidden symmetries of this spacetime following the above discussion.

\subsubsection{Conformal Killing vectors}

 By construction, our perturbed FLRW geometry admits fifteen CKVs which are the Lie-dragged deformations of the CKVs given in \Eq{CKVFLRW}. They can be written compactly as 
 \begin{align}
 \chi^{\mu}  \partial_{\mu} & = \mathcal{L}_\chi \bar{X}_\mu= \bar{\chi}^{\mu} \partial_{\mu} + \left( \xi^{\lambda} \nabla_{\lambda} \bar{\chi}^{\mu} - \bar{\chi}^{\lambda} \nabla_{\lambda} \xi^{\mu} \right)  \partial_{\mu}\, ,
 \end{align}
 which can be decomposed into 
 \beaq
 \label{Kil}
  \chi^{\mu} \partial_{\mu} & = \bar{T}^{\nu} \left( \delta^{\mu}{}_{\nu} - \nabla_{\nu} \xi^{\mu} \right) \partial_{\mu} \\ 
 & + \bar{\Omega}^{\nu}{}_{\rho} \left[  \delta^{\mu}{}_{\nu} \left( x^{\rho}+ \xi^{\rho}  \right) - x^{\rho} \nabla_{\nu} \xi^{\mu}\right] \partial_{\mu} \\
 & + \bar{\omega} \left[ x^{\mu} + \xi^{\mu} - x^{\rho} \nabla_{\rho} \xi^{\mu}\right] \partial_{\mu} \\
 & + \left[ 2 \bar{\beta}_{\rho} (x^{\rho} + \xi^{\rho})x^{\mu} +2\bar{\beta}_\rho x^\rho \xi^\mu- 2 \bar{\beta}_{\rho} x^{\rho} x^{\lambda} \nabla_{\lambda} \xi^{\mu} -\bar{ \beta}^{\mu} \eta_{\alpha\beta} x^{\alpha} (x^{\beta} + 2 \xi^{\beta}) + \bar{\beta}^{\lambda} (\eta_{\alpha\beta} x^{\alpha} x^{\beta}) \nabla_{\lambda} \xi^{\mu}\right] \partial_{\mu}
 \eeaq
 where $\bar{T}^{\mu}$, $\bar{\Omega}^{\nu}{}_{\rho}$, $\bar{\omega}$ and $\bar{\beta}_{\rho}$ are constant parameters respectively associated to the translations, rotations and boosts, dilatations and special-conformal transformations. We recall that the matrix $\bar{\Omega}^{\nu}{}_{\rho}$ is anti-symmetric. For completeness, we decompose them in terms of the components $(\xi^0, \xi^i)$ of the large diffeomorphism. The different generators  read 
 \begin{align}
 E^{\mu} \partial_{\mu}& =  \bar{T}^{0} \left\lbrace  \left[ 1- \etad{(\xi^0)} - \cH \xi^0 \right] \partial_\eta  - \left[\etad{(\xi^i)} + \cH \xi^i \right] \partial_i \right\rbrace , \\
 T^{\mu} \partial_{\mu} &  =  \bar{T}^i \left[ \left( \delta_i{}^j - \cH \xi^0 \delta_i{}^j - {\DB}_i \xi^j \right) \partial_j -\left( \partial_i \xi^0 + \cH {\gammaB}_{ij} \xi^j \right) \partial_\eta\right] ,\\
 R^{\mu} \partial_{\mu} & = \bar{\Omega}^i{}_j \left[ \left\{ (x^j + \xi^j ) \delta^k{}_i - x^j \left( {\DB}_i \xi^k + \cH \delta^k{}_i \xi^0\right)\right\} \partial_k - x^j \left( \partial_i \xi^0 + \cH {\gammaB}_{ij} \xi^j\right) \partial_\eta\right], \\
 B^{\mu} \partial_{\mu} & = \bar{\Omega}^0{}_i \left\lbrace \left[ (1- \etad{\xi}^0 - \cH \xi^0 ) x^i + \xi^i\right] \partial_{\eta} - x^i \left( \etad{\xi}^j + \cH \xi^j\right) \partial_j\right\rbrace, \\
  {\DB}^{\mu} \partial_{\mu} & = \bar{\omega} \left\lbrace \left[ \xi^0 + \eta \left( 1 - \etad{\xi}^0 - \cH \xi^0 \right) - x^i \left( \partial_i \xi^0 + \cH {\gammaB}_{ij} \xi^j  \right)\right] \partial_{\eta} \right. \nonumber \\
 & \;\;\;\;\;\; \left. + \left[ (1- \cH \xi^0) x^i + \xi^i - x^j {\DB}_j \xi^i -\eta \left(  \etad{(\xi^i)} + \cH \xi^i \right)\right] \partial_i \right\rbrace\, . 
 \end{align}
 By construction, these vector fields form the $\so(4,2)$ Lie algebra. Among them, only the spatial translations and rotations act as Killing vectors and are divergencefree. For the remaining nine generators that are constructed from the CKVs of the flat FLRW metric, they act as CKVs for the perturbed FLRW geometry such that
 \begin{align}
 \cL_{\chi} g_{\mu\nu} = \Omega g_{\mu\nu} \qquad \text{with} \qquad \Omega = \bar{\Omega} +  \cL_{\xi} \bar{\Omega}
 \end{align}
 where $\bar{\Omega}$ is the divergence of the background CKV. One can check that these CKVs indeed generate exact conserved charges for the geodesic motion on the perturbed FLRW geometry, as expected, allowing one to integrate the geodesic equation. 
 
 At this stage, it is important to realize that the Lie-dragged isometries do not share the same global status as their background seeds. Indeed, because the generator of the large-gauge transformation $\xi^{\mu} \partial_{\mu}$ admits free time-dependent functions, through the frame freedom encoded in the  function $m(\eta)$, the vectors $c^i(\eta)$ and $g_i(\eta)$ and the tensor $\Omega_{ij}(\eta)$, one obtains one $\so(4,2)$ Killing algebra for each choice of frame.  Let us now turn to the generators of the hidden symmetries of the soft perturbations.

\subsubsection{Killing-Yano and Killing tensors}

As discussed in the previous section, the hidden symmetries are generated by the Killing-Yano and Killing tensors. We have seen that the FLRW background possesses twenty two Killing tensors and four KY tensors. Since all the background KTs are reducible, we shall only focus on the KY two-forms. 

As shown above, the Lie-dragged background KY tensors are exact solutions of the KY equation on the perturbed FLRW geometry. 
The deformed KY tensors take the following compact expressions
\begin{align}
Y_{\mu\nu} & = \bar{Y}_{\mu\nu} + \cL_{\xi} \bar{Y}_{\mu\nu}  \\
& = \bar{Y}_{\mu\nu} + \xi^{\lambda} \partial_{\lambda} \bar{Y}_{\mu\nu} + \bar{Y}_{\mu\lambda} \partial_{\nu} \xi^{\lambda} + \bar{Y}_{\lambda\nu} \partial_{\mu} \xi^{\lambda}\, .
\end{align}
Using the form of the background KY tensor, $\bar{Y}_{ij} \rd x^i \wedge \rd x^j$, given in \Eq{KKY}, the two non-vanishing spatial components are easily computed and read
\begin{align}
 \label{KY1}
Y_{ij}^{(I)} 
& =  (1+ 3 \cH \xi^0)\bar{Y}_{ij}  + \bar{Y}_{ik} \partial_{j} \xi^{k} + \bar{Y}_{k j} \partial_{i} \xi^{k} \\
Y_{ij}^{(4)} 
& =  (1+ 3 \cH \xi^0)\bar{Y}_{ij} + a^3 \epsilon_{ijk}\xi^k  + \bar{Y}_{ik} \partial_{j} \xi^{k} + \bar{Y}_{k j} \partial_{i} \xi^{k} 
\end{align}
while the off-diagonal component reads
\begin{align}
\label{last}
Y_{0i}  & = \bar{Y}_{k i} \etad{\xi}^{k}  = \bar{Y}_{ki} {\gammaB}^{kj} {\DB}_j  \xi^0
\end{align}
where we have used the gauge condition~\eqref{shift} in the last equality and the fact that $\bar{Y}_{0i} =0$. Since we have four distinct background KY tensors given by \Eq{KKY}, the perturbed FLRW geometry inherits four deformed KY tensors.  
As already mentioned, these KY tensors can be used for different purposes. First, they provide the building blocks needed to derive conserved charges for the motion of spinning particle in the perturbed FLRW geometry. Second, and most importantly, they can be used to construct conserved currents for the gravitational field itself. These charges introduced initially by Penrose for the vacuum case and generalized by Kastor and Traschen to the non-vacuum case are not the standard ones \cite{Penrose:1982wp, Kastor:2004jk} as they are built not from the Killing vectors but from the Killing-Yano tensors of the geometry.

However, the main goal of this last section was to emphasize and explicitly derive the Killing symmetries in the presence of LWMs in cosmology. Exploring the different charges associated to these Killing symmetries goes beyond the scope of this article. A more detailed study of these charges will be presented elsewhere. We can now summarize our results.

\section{Discussion}

\label{C}

In this work we have investigated two sets of symmetries of linear cosmological perturbations. First, we have reinvestigated the soft symmetries of the linearized Einstein's equations for the cosmological perturbations within the Newtonian gauge. These symmetries stand as symmetries of the theory, \ie~General Relativity coupled to a perturbed perfect fluid at the linear order. Second, we have discussed the explicit and hidden Killing symmetries of the perturbed FLRW geometries where the inhomogeneity corresponds to a LWM. These Killing symmetries stand as symmetries of a given solution. 

Let us stress that soft modes have also been considered in the context of infrared divergences, see Ref.~\cite{Tanaka:2013caa} for a review. There, residual gauge transformations are studied because they generate apparent infrared contributions that should not affect genuinely observable quantities inside a local observable region. The emphasis is on identifying and removing gauge artefacts from correlation functions and loop corrections. In the approach followed in the present work, the emphasis is instead on the symmetry structure of cosmological perturbation theory viewed from the perspective of large-gauge transformations and boundary degrees of freedom. The criterion adopted here is whether a residual diffeomorphism modifies the geometry of the cosmological patch and its quasi-local observables (energy, angular momentum, etc.), rather than whether it can be smoothly continued to finite momentum or eliminated from local correlators.

Let us summarize our key results. 
\begin{itemize}
\item \textit{Extended local soft symmetries:} Building on previous works, we have demonstrated that the LGTs w.r.t. the Newtonian gauge are infinite-dimensional and local. The vector fields generating these symmetries depend on free time-dependent functions that appear in the low-multipole sector of the multipole decomposition of its components. Their expressions are given by \Eqs{TDifff} and~\eqref{SDifff}. This result contrasts with previous works where the soft symmetries were found to be global \cite{Weinberg:2003sw, Creminelli:2013mca, Pajer:2017hmb}. In these studies, a soft mode is said to be physical if it solves the linearized Einstein equations away from the zero-momenta limit in Fourier space, \ie in the limit $k\rightarrow0$. We argued that this definition lacks a covariant formulation and should thus be used with care. Using instead the standard approach that consists in declaring a soft mode physical when it affects the boundary observable of the system, we have argued that the whole tower of low-multipole soft modes discarded in previous studies have to be considered as bona-fide physical modes. Since these modes live in the kernel of the spatial gradients appearing in the Einstein equations, their time-dependence is unconstrained, giving rise to local soft symmetries. The upgrade of the soft cosmological symmetries from a set of global to a set of local symmetries is the first result of this work.   
\item \textit{Effects on the energy and angular-momentum of the cosmological patch:} In \Sec{D}, we have reviewed the definition of the mean-curvature vector (\ref{MCV}), its dual (\ref{MCVD}) and the associated twist vector. We have then discussed the notion of mean-curvature energy (\ref{EN}) and angular momentum (\ref{MCVL}), and reviewed why they provide a well-adapted notion of energy of dynamical spacetimes such as cosmological geometries. Using these tools, we have shown (i) how the monopole $m$ and dipole $g_i x^i$ soft modes  modify the quasi-local energy of the cosmological patch and the position of the cosmological horizon and (ii) how the dipole soft mode $\Omega^i{}_j x^j$ introduces a non-vanishing angular momentum and thus an axis of rotation in the patch. This parallels the well-known effect of monopole and (axial) dipole in black-hole perturbation theory, which shift respectively the mass and the spin of the black hole in the linear regime. While a more systematic investigation of the boundary effects of each soft mode would be desirable, the results presented here provide a first concrete example of how these low-multipole soft modes affect the quasi-local properties of the cosmological patch. 
\item \textit{Physical interpretation of the low-multipole ambiguity:} As mentioned in the introduction, the existence of local soft symmetries implies that the Einstein's equations allow for the existence of the unconstrained low-multipole modes \eqref{SM0}-\eqref{SM1}. We have argued that this freedom should be understood in terms of a frame ambiguity, similar to the BMS frame ambiguity known in asymptotically-flat spacetimes. Properly understanding how this ambiguity affects the computation of cosmological observables within cosmological perturbation theory begs for more work. Let us however note that since low-multipole modes affect the geodesic motion of test probes, as shown in \Sec{B}, the trajectory of photons should be sensitive to their presence, i.e.~the low-multipole soft modes ambiguity discussed in this work could be reflected in the low-multipole sector of the temperature anisotropies of the CMB. It is tempting to wonder whether this could alleviate the CMB low-multipole anomaly~\cite{Copi:2010na, Patel:2024oyj, Jung:2024slj}. It would also be interesting to explore whether the time-dependence of these soft modes could be constrained by imposing suitable energy conditions.

\item \textit{Algebraically-special cosmological perturbations and the Killing symmetries: }In the second part of this work, we have further explained how the adiabatic modes preserve the Killing symmetries of the FLRW background. While this is intuitively what one expects for these soft modes, the results presented in the second section allow one to make this point concrete. Indeed, LWMs on top of the FLRW background are commonly understood as soft deformations of the cosmological patch that can be reabsorbed into a new background geometry. From that point of view, it appears natural that their presence only deforms, but does not break, the FLRW properties. The machinery of the LGTs allows us to make this point explicit. By construction, the LWMs being built from a diffeomorphism of the flat FLRW geometry, they inherit all the properties of the FLRW geometry through the action of Lie-dragging. A direct consequence is that the Weyl tensor remains vanishing in the presence of these LWMs, such that these adiabatic modes stand as algebraically special cosmological perturbations of Petrov type O. To our knowledge, this point has not been properly appreciated and could be relevant to the separate-universe approach, and the related stochastic-inflation formalism. As we have shown, a second consequence is that these large-scale inhomogeneous cosmological geometries enjoy the same (conformal) isometry group as FLRW, \ie the SO$(4,2)$ group. The expression of the (conformal) Killing vectors is given by \Eq{Kil}. This is also the case for the four Killing-Yano 2-forms of the FLRW symmetry, which provide four KY tensors given by \Eqs{KY1}-\eqref{last} that survive in the presence of the LWMs. The fact that these explicit and hidden Killing symmetries still hold for LWMs, while being a simple consequence of covariance, opens a path to analytical insight into key questions such as the effect of non-trivial large scalar perturbations on the propagation of light or gravitational waves. 
\item \textit{Refinement of the notion of the FLRW model and the cosmological patch}: The soft and Killing symmetries discussed in this work naturally relax the notion of what one considers as being an FLRW spacetime geometry. Indeed, since any FLRW background dressed with a given LWM shares the same Killing symmetries and Petrov type with the flat FLRW background, the isometries can be used to characterize this geometry. However, by construction, the soft symmetries provide the right tool to distinguish between all these FLRW-like geometries as they all differ by their boundary properties. Just as in the AFS context, the best concept to grasp the structure of the geometry of the cosmological patch is the notion of degenerate cosmological vacua. Concretely, the flat FLRW without LWMs can be considered as the true cosmological vacuum, on top of which one can construct an infinite set of excitations manifesting through long-wavelength deformations of the patch.
\end{itemize} 
These results provide a preliminary step into the interplay between the computation of cosmological observables and the geometrical description of the cosmological patch. The natural extensions of the present work can be organized as follows.  

First, the identification of the extended LGTs presented in this work needs to be complemented with a computation of the associated flux-balance laws, which encode the dynamics of the cosmological patch. This step is crucial to understand how the boundary observables characterizing the cosmological patch are modified by the action of one of the LGTs identified in this work. This should allow one to clarify the long-standing debate on the effect of LWMs on various cosmological observables, which are computed in the bulk geometry. Indeed, while the LWMs can indeed be added or removed locally by the appropriate change of coordinates, this induces a change in the boundary and thus on the gravitational observables that label the geometry of the patch, as demonstrated by the computation of the energy and the angular momentum of the patch in \Sec{D}. For instance, the presence of a time-dependent dipole introduces a symmetry axis into the cosmological patch. A clear understanding of this deformation of the geometry of the patch requires a computation of the associated flux-balance law. This task requires to fully take into account the role of the boundary in the computation of cosmological observables, an aspect which has remained largely ignored in the investigation of the role of LWMs in cosmology. This will be presented in a forthcoming work \cite{NEXT}.

 A second perspective opened by the Killing symmetries discussed in \Sec{B} is to provide conserved charges to integrate the motion of test fields and test particles in the presence of non-trivial LWMs. Consider for instance the flat FLRW background with the large-scale anisotropy given by \Eq{ANI}. At first glance, this non-trivial cosmological geometry would not be recognized as being invariant under the conformal group, nor having four Carter-like constants, \ie Killing tensors that descend from the existence of Killing-Yano tensors. Nevertheless, the existence of this set of symmetries is guaranteed by the covariance of the construction and allows one to capture the effects of such anisotropy on the propagation of test particles or fields. For instance, applying this approach to revisit gravitational-lensing computations could provide new analytical results relevant for interpreting the contribution of large-scale perturbations in cosmological data.

 Let us finally note that while our analysis is restricted to linear perturbation theory, it would be interesting to determine whether the low-multipole ambiguity survives at nonlinear order. In particular, the requirement mentioned below Eq.~(\ref{EOM7}), namely that the gauge condition remains stable under the Lie bracket, might not be fulfilled at quadratic order. That could prevent the soft symmetries from forming a closed algebra, and may require to use another expansion scheme than the multipolar decomposition (the case of asymptotically-flat and radiative spacetime, treated at the fully non-linear level, provides one such example~\cite{Fiorucci:2021pha, Campiglia:2020qvc, Flanagan:2015pxa, Compere:2020lrt, Compere:2018ylh, Freidel:2021fxf, Geiller:2022vto, Geiller:2024amx}).

\acknowledgments
J. Ben Achour is grateful to Sk. J. Hoque and A. Fiorucci for enlightening discussions on the notion of large-gauge transformations and the covariant phase-space formalism. We also thank B. Blachier, C. Ringeval and M. A. Gorji for useful comments and interesting discussions.

\newpage
\appendix

\section{Christoffel symbols for cosmological perturbations}
 Let us start by decomposing the metric and write its inverse. We can compactly write
 \be
 \rd s^2 = g_{\mu\nu} \rd x^{\mu} \rd x^{\nu} = a^2(\eta) \left( \eta_{\mu\nu} + h_{\mu\nu}\right) \rd x^{\mu} \rd x^{\nu}
 \ee
 where $\eta_{\mu\nu}$ is the Minkowski metric and $h_{\mu\nu}$ contains the rescaled perturbations. Working at this level, the inverse metric is given by
 \be
 g^{\mu\nu} = \frac{1}{a^2(\eta)} \left( \eta^{\mu\nu} - h^{\mu\nu}\right) + \cO(h^2)\, .
 \ee
 Now we can decompose the rescaled perturbation as follows
 \be
 h_{\mu\nu} = \mat{cccc}{- 2 \Phi  & - N_i \\ - N_j & - 2 \Psi \delta_{ij} + 2 E_{ij}} \;, \qquad h^{\mu\nu} = \mat{cccc}{- 2 \Phi & + N^i \\ + N^j & - 2 \Psi \delta^{ij} + 2 E^{ij}} 
 \ee
 where we have used that $h^{\mu\nu} = \eta^{\mu\rho} h^{\nu\sigma} h_{\rho\sigma}$ to compute the inverse form. Since $\eta^{00} =-1$ and $\eta^{ij} = \delta^{ij}$, only the cross term changes sign. It follows that the full metric reads
 \begin{align}
 g_{\mu\nu} = a^2(\eta) \mat{cccc}{- 1 - 2\Phi & - N_i \\ - N_i &  \left( 1-2\Psi\right) \delta_{ij} + 2 E_{ij}} \;, \qquad  g^{\mu\nu} = \frac{1}{a^2(\eta)} \mat{cccc}{- 1 +2\Phi & - N^i \\ - N^i &  \left( 1+2\Psi \right) \delta^{ij} - 2 E^{ij}} .
 \end{align}
In the end, we obtain the following metric of the perturbed flat FLRW universe
\begin{align}
\rd s^2 & = g_{\mu\nu} \rd x^{\mu} \rd x^{\nu} \\
& = a^2(\eta) \left\{ - \left( 1+ 2\Phi \right) \rd \eta^2  +2 N_i \rd \eta \rd x^i + \left[ \left( 1-2\Psi \right) \delta_{ij} + 2 h_{ij}\right] \rd x^i \rd x^j \right\}
\end{align}
where the scalar fields $(\Phi,\Psi )$, the vector $B_i$ and the tensor $E_{ij}$ stand as the perturbations. 
Notice that one can further decompose the perturbations into scalar, vector and tensor contributions as follows. One has
\begin{align}
N_{i} & =  \partial_i B + B_i\, , \\
h_{ij} & =  \partial_i \partial_j E + \frac{1}{2}  \partial_{(i} E_{j)}  + E_{ij}\, .
\end{align}
In this section, we provide the different ingredients to write down and study the Killing equation for this spacetime. The Christoffel symbols read
 \begin{align}
 \Gamma^0{}_{00} & =  \cH + \etad{\Phi} \\
 \Gamma^0{}_{0i} & =  \partial_i \Phi  \\
 \Gamma^0{}_{ij} &  = \cH \left[ \left( 1- 2 \Phi - 2 \Psi \right) {\gammaB}_{ij} + 2 E_{ij}\right] - {\gammaB}_{ij} \etad{\Psi} + \etad{E}_{ij}+ \frac{1}{2} \left( \partial_i B_j + \partial_j B_i\right) 
 \end{align}
 while the second part is given by
  \begin{align}
 \Gamma^i{}_{00} &  = {\gammaB}^{ij} \partial_j \Phi - \cH B^i - \etad{B}^i  \\
 \Gamma^{i}{}_{0j} & =  \frac{1}{2} {\gammaB}^{ik} \left( {\DB}_k B_j - {\DB}_j B_k \right) + \cH \left[ {\gammaB}^i{}_j - 2 E^{ik} {\gammaB}_{kj} \right] + {\gammaB}^{ik} \etad{E}_{kj} -  \etad{\Psi} \delta^i{}_j \\
 \Gamma^i{}_{jk}  & =  {\gammaB}^k{}_{ij}  + \frac{1}{2} \left( {\DB}_i E^k{}_{j} + {\DB}_j E^k{}_{i} - {\DB}^k E_{ij}\right) - (\delta^k{}_j\partial_i \Psi + \delta^k{}_i \partial_j \Psi - {\gammaB}_{ij} \partial^k \Psi  )
 \end{align}
and it follows that
\begin{align}
\Gamma^k{}_{ki} = {\gammaB}^k{}_{ki} - 3 \partial_i \Psi 
\end{align}
where we have used that $E^k{}_k =0$ and ${\DB}_i E^i{}_j =0$.
This provides the required ingredients for computing further tensorial quantities.

\section{Linearized Einstein equations for the soft symmetry generators}

\label{FE}
In the Newtonian gauge, the linearized Einstein equations are given by
\begin{align}
\label{eq:perturbedEinstein:scalar:Newton:in:2}
 \Delta \Psi  -3\cH (\etad{\Psi}+ \cH\Phi)  & = \frac{\kappa}{2} a^2 \delta \rho\, ,  \\
 \label{eq:perturbedEinstein:mom:Newton:in}
 {\DB}_i ( \etad{\Psi} + \cH \Phi  ) & = -  \frac{ \kappa}{2} a^2 \rho (1+\omega) {\DB}_i v \, ,\\
 \label{eq:perturbedEinstein:press:Newton:in}
   \etadd{\Psi } + \cH (\Phi'+2 \etad{\Psi})  + (2\etad{\cH} + \cH^2) \Phi - \frac{1}{3} \Delta(\Psi - \Phi)  & =  \frac{1}{2}  \kappa a^2 \delta P \, , \\
   \label{eq:perturbedEinstein:ani:Newton:in}
 \Delta_{ij} (\Psi - \Phi) & =  \kappa a^2 P \Delta_{ij}  \pi \, ,
\end{align}
in the scalar sector, while the vector and tensor perturbations satisfy the following dynamical equations
\begin{align}
\label{eq:vector:eom:2:app}
D_{(i}\left[\Phi_{j)}'+2\mathcal{H}\Phi_{j)}\right]& =\kappa P a^2 D_{(i}\pi_{j)} \\
\label{eq:vector:eom:1:app}
\Delta\Phi_i + 4 \left(\cH'-\cH^2\right)\Phi_i & = -2\kappa a^2 \rho (1+\omega)v_i  \, ,\\
\label{eq:tensor:eom:app}
 \etadd{E_{ij}} + 2 \cH \etad{E_{ij}}  - \Delta E_{ij} = & \kappa a^2 P \pi_{ij}\, .
\end{align}
Here, the anisotropic stress and the non-adiabatic perturbation can be dropped since they are not generated by the soft diffeomorphisms as discussed around \Eq{eq:matter:pert:xi}.
\begin{itemize}
\item Let us first consider the scalar constraint, \ie \Eq{eq:perturbedEinstein:scalar:Newton:in:2}. In terms of the diffeomorphism component, \ie making use of \Eq{pot1}, \eqref{pot2} and \eqref{eq:matter:pert:xi}, one has
\begin{align}
& {\DeltaB} \Psi - 3 \cH (\etad{\Psi} + \cH \Phi) - \frac{\kappa}{2} a^2 \delta \rho \nn  \\
& = -   {\DeltaB} ( \cH \xi^0 + \frac{1}{3} {\DB}_i \xi^i) + 3 \cH \left[  \etad{\cH} \xi^0 + \frac{1}{3}D_i\etad{(\xi^i)}- \cH^2 \xi^0 \right] - \frac{\kappa}{2} a^2 \xi^0 \etad{\rho} \, .
\end{align}

From \Eq{shift}, one has $\DB_i\etad{(\xi^i)}=\DeltaB\xi^0$, hence the above reduces to
\begin{align}
& {\DeltaB} \Psi - 3 \cH (\etad{\Psi} + \cH \Phi) - \frac{\kappa}{2} a^2 \delta \rho   = -    \frac{1}{3}  {\DeltaB} {\DB}_i \xi^i +  \left( \Ji{3 \cH \etad{\cH}  -3 \cH^3 - \frac{\kappa}{2} a^2 \etad{\rho}} \right)\xi^0  = 0\, .
\end{align}
Making use of the background equations~\eqref{eq:Friedmann:1}-\eqref{eq:conservation}, one can show that the second term vanishes, which leads to 
\begin{align}
& {\DeltaB} \Psi - 3 \cH (\etad{\Psi} + \cH \Phi) - \frac{\kappa}{2} a^2 \delta \rho   = -    \frac{1}{3}  {\DeltaB} {\DB}_i \xi^i  =0\, .
\end{align}
The scalar constraint~\eqref{eq:perturbedEinstein:scalar:Newton:in:2} thus requires that  ${\DeltaB} {\DB}_i \xi^i=0$.
\item Let us consider the momentum constraint~\eqref{eq:perturbedEinstein:mom:Newton:in},
\begin{align}
{\DB}_i \left[ \etad{\Psi} + \cH \Phi + \frac{\kappa}{2} a^2 \rho (1+\omega) v \right] & = {\DB}_i \left\lbrace \left[\Ji{ \cH^2-\etad{\cH}-\frac{\kappa}{2}a^2\rho(1+\omega)}\right]\xi^0 - \frac{1}{3} \DB_k\etad{(\xi^k)} \right\rbrace\qquad  \;\;\;\\
& = - \frac{1}{3}  {\DB}_i \DB_k\etad{(\xi^k)}
  = - \frac{1}{3} {\DeltaB} {\DB}_i \xi^0=0\, .
\end{align}
Here again, the second term vanishes if the background equations~\eqref{eq:Friedmann:1}-\eqref{eq:conservation} are enforced, and from \Eq{shift}, one has $\DB_i\etad{(\xi^i)}=\DeltaB\xi^0$. The momentum constraint equation thus gives ${\DeltaB} {\DB}_i \xi^0=0$.
\item The trace of the dynamical equation~\eqref{eq:perturbedEinstein:press:Newton:in} leads to
\begin{align}
&    \etadd{\Psi } + \cH (\Phi'+2 \etad{\Psi})  + (2\etad{\cH} + \cH^2) \Phi - \frac{1}{3} \Delta(\Psi - \Phi)  - \frac{1}{2}  \kappa a^2 \delta P \nonumber \\
& = \xi^0 \left(\Ji{  -\etadd{\cH}  +  \cH^3 + \cH \etad{\cH} - \frac{\kappa}{2} a^2 \etad{P} }\right) - \cH \etadd{\xi}^0 - 2\etad{\cH} \etad{\xi}^0 - 2 \cH^2 \etad{\xi}^0 + (2\etad{\cH} + \cH^2) \etad{\xi}^0 + \cH \etadd{\xi}^0 + \cH^2 \etad{\xi}^0 \nn \\
& + \frac{1}{3} \left[ \frac{1}{3} \Delta D_i \xi^i - (D_i \xi^i)'' - 2 \cH (D_i \xi^i)' \right] + \frac{1}{3} \Delta \left[(\xi^0)'  + 2 \cH \xi^0\right] .
\end{align}
Upon imposing the background equations, the colored terms cancel out while the rest of the second line obviously cancels out. Further using ${\DeltaB} {\DB}_i \xi^0=0$ from the momentum constraint, one arrives at
\begin{align}
&    \etadd{\Psi } + \cH (\Phi'+2 \etad{\Psi})  + (2\etad{\cH} + \cH^2) \Phi - \frac{1}{3} \Delta(\Psi - \Phi)  - \frac{1}{2}  \kappa a^2 \delta P \\
& =  \frac{1}{3} \left[  - (D_i \xi^i)'' - 2 \cH (D_i \xi^i)' \right] + \frac{1}{3} \Delta \left[(\xi^0)'  + 2 \cH \xi^0\right]\, ,
\end{align}
which cancels out when using \Eq{shift}. Therefore, the dynamical equation~\eqref{eq:perturbedEinstein:press:Newton:in} is automatically satisfied and does not impose further conditions.
\item Finally, the remaining traceless equation~\eqref{eq:perturbedEinstein:ani:Newton:in} leads to
\begin{align}
\Delta_{ij} ( \Phi - \Psi) & = \left({\DB}_{(i} {\DB}_{j)} - \frac{2}{3} {\gammaB}_{ij} {\DeltaB}\right) \left[ \etad{(\xi^0)}  + 2 \cH \xi^0 + \frac{1}{3} {\DB}_k \xi^k\right] =0\, .
\end{align}
\end{itemize}
We can now turn to the remaining equations involving the tensor and vector perturbations.
\begin{itemize}
\item
In the tensor sector, \Eq{eq:tensor:eom:app} leads to
\begin{align}
 \etadd{E_{ij}} + 2 \cH \etad{E_{ij}}  - \Delta E_{ij} &  = \frac{1}{2} \left[{\gammaB}_{jk} {\DB}_i  + {\gammaB}_{ik} {\DB}_j  - \frac{2}{3}  {\gammaB}_{ij} {\DB}_k  \right] \left[ (\xi^k)'' + 2\cH (\xi^k)'\right] \nn \\
& - \frac{1}{2} \Delta \left[{\gammaB}_{jk} {\DB}_i \xi^k + {\gammaB}_{ik} {\DB}_j \xi^k - \frac{2}{3}  {\gammaB}_{ij} {\DB}_k \xi^k \right] = 0\, .
\end{align}
\item In the vector sector, using that $v_i = \Phi_i$, \Eq{eq:vector:eom:1:app} reads 
\begin{align}
4\left[ \Ji{\cH'-\cH^2+ \frac{ \kappa}{2} a^2 \rho (1+\omega) }+\frac{1}{4} \Delta \right]\Phi_i & =   \Delta \Phi_i =
 \Delta \left[ {\DB}_i \xi^0 - {\gammaB}_{ij} \etad{(\xi^j)}\right] =0 
   \end{align}
where the colored terms cancel out when using the background equations.
\item Finally, \Eq{eq:vector:eom:2:app} leads to
\begin{align}
D_{(i}\left[\Phi_{j)}'+2\mathcal{H}\Phi_{j)}\right] = & D_{(i}\left[{\DB}_{j)} (\xi^0)'+2\cH  {\DB}_{j)} \xi^0 - \gamma_{j)k}  (\xi^k)'' - 2 \cH \gamma_{j)k}  (\xi^k)'\right]\\
=& \frac{1}{a^2} \left\lbrace  a^2D_{(i}\left[{\DB}_{j)} \xi^0 - \gamma_{j)k}  (\xi^k)' \right] \right\rbrace ' =0\, . 
\end{align}
We thus find that ${\DB}_i \xi^0 - {\gammaB}_{ij} \etad{(\xi^j)}$ is annihilated by two second-order differential operators.
\end{itemize}

\section{Lie derivative and Lie dragging}

\subsection{Commutator between the Lie and covariant derivatives}
\label{app:commutator:Lie:covariant:derivative}

In this appendix, we derive the commutator between the Lie derivative and the covariant derivative. It corresponds to the variation of the Christoffel symbol under the action of the Lie derivative.
The commutator between covariant derivatives can be expressed in terms of the Riemann tensor
\begin{align}
\left[ \nabla_{\mu} \nabla_{\rho}  - \nabla_{\rho} \nabla_{\mu} \right] \xi^{\alpha} =  R^{\alpha}{}_{\sigma \mu \rho} \xi^{\sigma}
\end{align}
where
\begin{align}
R^{\rho}{}_{\beta\gamma\delta} = \partial_{\gamma} \Gamma^{\rho}{}_{\beta\delta} - \partial_{\delta} \Gamma^{\rho}{}_{\beta\gamma} + \Gamma^{\rho}{}_{\mu\gamma}  \Gamma^{\mu}{}_{\beta\delta} - \Gamma^{\rho}{}_{\mu\delta} \Gamma^{\mu}{}_{\beta\gamma}\, .
\end{align}
Consider the variation under a diffeomorphism of the covariant derivative of a covector. It reads
\be
\delta \left( \nabla_{\mu} X_{\nu} \right) -  \nabla_{\mu} \delta X_{\nu} = -  \delta \Gamma^{\rho}{}_{\mu\nu} X_{\rho}\, .
\ee
Now, we can show that for a given covector, one has
\begin{align}
\cL_{\xi} \left( \nabla_{\mu} X_{\nu}\right) - \nabla_{\mu} \left( \cL_{\xi} X_{\nu}\right) = - X_{\lambda} \left[ \nabla_{\mu} \nabla_{\nu} \xi^{\lambda} + R^{\lambda}{}_{\nu\rho\mu} \xi^{\rho}\right] .
\end{align}
Interestingly, one finds that the commutator is proportional to the integrability condition for the Killing equation. Concretely, it implies that the commutator vanishes only when the diffeomorphism $\xi^{\alpha} \partial_{\alpha}$ is a (conformal) Killing vector. 
In the end, we obtain the following key result
\be
\label{KFA}
\boxed{\delta \Gamma^{\lambda}{}_{\mu\nu} = \nabla_{\mu} \nabla_{\nu} \xi^{\lambda} + R^{\lambda}{}_{\nu\rho\mu} \xi^{\rho}\, .}
\ee
Let us now show explicit use of this formula.

\subsection{Commutation between Lie derivative and Killing-Yano tensor equation}
\label{KYapp}

In this appendix, we provide the proof of \Eq{KYCOM} using the formula (\ref{KFA}). Consider the perturbed metric 
\be
g_{\mu\nu} = \cL_{\chi} \bar{g}_{\mu\nu}
\ee
and let us assume that $\bar{g}$ admits a KY two form $\bar{Y}_{\nu\alpha}$ defined by $\bar{\nabla}_{(\mu} \bar{Y}_{\nu)\alpha} = 0$. Then, one has 
\begin{align}
\nabla_{\mu} Y_{\nu\alpha} & = \bar{\nabla}_{\mu} \left( \chi^{\lambda} \nabla_{\lambda} \bar{Y}_{\nu\alpha} + \bar{Y}_{\nu\lambda} \nabla_{\alpha} \chi^{\lambda} +  \bar{Y}_{\nu\lambda} \nabla_{\alpha} \chi^{\lambda} \right) - \delta_{\chi} \Gamma^{\rho}{}_{\mu\nu} \bar{Y}_{\rho\alpha} - \delta_{\chi} \Gamma^{\rho}{}_{\mu\alpha} \bar{Y}_{\nu\rho} \nonumber \\
& =   \bar{\nabla}_{\mu}  \chi^{\lambda} \bar{\nabla}_{\lambda} \bar{Y}_{\nu\alpha} + \chi^{\lambda} \bar{\nabla}_{\mu} \bar{\nabla}_{\lambda} \bar{Y}_{\nu\alpha} + \bar{\nabla}_{\mu} \bar{Y}_{\nu\lambda} \bar{\nabla}_{\alpha} \chi^{\lambda} + \cancel{\bar{Y}_{\nu\lambda} \bar{\nabla}_{\mu} \bar{\nabla}_{\alpha} \xi^{\lambda} } + \bar{\nabla}_{\mu} \bar{Y}_{\lambda \alpha} \bar{\nabla}_{\nu} \chi^{\lambda} + \cancel{\cancel{\bar{Y}_{\lambda \alpha} \bar{\nabla}_{\mu} \bar{\nabla}_{\nu} \chi^{\lambda} }} \nn \\
& \;\;\; - \cancel{\bar{Y}_{\rho\alpha} \bar{\nabla}_{\mu} \bar{\nabla}_{\nu} \chi^{\rho}} - \cancel{\cancel{\bar{Y}_{\nu\rho} \bar{\nabla}_{\mu} \bar{\nabla}_{\alpha} \chi^{\rho}}} - \Ji{ \chi^{\lambda} \left( \bar{R}^{\rho}{}_{\nu\lambda\mu} \bar{Y}_{\rho\alpha} + \bar{R}^{\rho}{}_{\alpha\lambda\mu} \bar{Y}_{\nu\rho}\right) } \nn \\
& = \cancel{\chi^{\lambda} \bar{\nabla}_{\mu} \bar{\nabla}_{\lambda} \bar{Y}_{\nu\alpha} } - \Ji{ \cancel{\chi^{\lambda} \bar{\nabla}_{\mu} \bar{\nabla}_{\lambda} \bar{Y}_{\nu\alpha} }+ \chi^{\lambda} \bar{\nabla}_{\lambda} \bar{\nabla}_{\mu} \bar{Y}_{\nu\alpha} }+  \bar{\nabla}_{\mu}  \chi^{\lambda} \bar{\nabla}_{\lambda} \bar{Y}_{\nu\alpha}  +  \bar{\nabla}_{\mu} \bar{Y}_{\nu\lambda} \bar{\nabla}_{\alpha} \chi^{\lambda}  +  \bar{\nabla}_{\mu} \bar{Y}_{\lambda \alpha} \bar{\nabla}_{\nu} \chi^{\lambda} \nn \\
& = \cL_{\chi} \left( \bar{\nabla}_{\mu} \bar{Y}_{\nu\alpha} \right) 
\end{align}
where the color terms in the third expression give rise to the colored terms in the fourth one.
Notice that we never used any properties of the KY tensor so far, so that the above applies to any rank-2 tensor. Then, it is direct to see that by construction, $Y_{\nu\alpha} =  \cL_{\chi} \bar{Y}_{\mu\nu}$ will be a KY tensor w.r.t. $g = \cL_{\chi} \bar{g}_{\mu\nu}$ if $\bar{Y}_{\nu\alpha}$ is a KY w.r.t. $\bar{g}$. Indeed, one has 
\begin{align}
\nabla_{\mu} Y_{\nu\alpha} + \nabla_{\nu} Y_{\mu\alpha} = \cL_{\chi} \left( \bar{\nabla}_{\mu} \bar{Y}_{\nu\alpha} + \bar{\nabla}_{\nu} \bar{Y}_{\mu\alpha} \right) = 0.
\end{align}
It is direct to see that it automatically generalizes to any Killing symmetries of the background.

\subsection{Lie-dragging Killing vector fields}

\label{CKVapp}

Consider a manifold $(\bar{M}, \bar{g})$ which admits a conformal Killing vector $\bar{Y}^{\mu} \partial_{\mu}$ such that it satisfies
\be
\cL_{\bar{Y}} \bar{g} =  \bar{\Omega} \bar{g}\, .
\ee
Now, let us construct a new inequivalent solution $(M, g)$ by a soft diffeomorphism $\chi^{\mu} \partial_{\mu}$ such that the new metric is given by
\be
g = \delta \bar{g} = \cL_{\xi} \bar{g}
\ee
and let us consider the vector field $\chi^{\mu} \partial_{\mu}$ in a perturbative expansion such that $\chi^{\mu} \partial_{\mu} = \partial_{\mu} + \epsilon \xi^{\mu} \partial_{\mu}$ where $\epsilon$ keeps track of the order of the perturbation. Then, the new metric decomposes into 
\be
g = \delta \bar{g} = \cL_{\chi} \bar{g} = \bar{g} +\epsilon  \cL_{\xi} \, .\bar{g}
\ee
Let us now consider the Lie-dragged Killing vector, which reads
\be
Y = \cL_{\chi} \bar{Y} = \bar{Y} + \epsilon \cL_{\xi} \bar{Y} = \bar{Y} + \epsilon [\xi, \bar{Y}]\, .
\ee
By construction, this Lie-dragged conformal Killing vector is a conformal Killing vector for the perturbed metric $g$ up to first order in the perturbations. Indeed, it is direct to see that
\begin{align}
\cL_Y g & = \cL_{\bar{Y}} g + \epsilon \cL_{[\xi ,\bar{Y}]} g \\
 & =  \bar{\Omega} \bar{g} +  \epsilon  \cL_{\bar{Y}} \cL_{\xi} g +  \epsilon  \cL_{[\xi, Y]} \left( \bar{g} + \epsilon \cL_{\xi} \bar{g}\right) \\
& =  \bar{\Omega} \bar{g} +  \epsilon \left[  \cL_{\bar{Y}} \cL_{\xi} g +  \cL_{[\xi, Y]}  \bar{g} \right]  + \cO(\epsilon^2) \\
& =  \bar{\Omega} \bar{g} +  \epsilon \cL_{\xi} \left(  \cL_{\bar{Y}} \bar{g} \right) + \cO(\epsilon^2) \\
& =  \bar{\Omega} \bar{g} +  \epsilon \cL_{\xi} \left(  \bar{\Omega} \bar{g} \right) + \cO(\epsilon^2)  \\
& =  \left(  \bar{\Omega} + \epsilon \cL_{\xi}  \bar{\Omega}\right)  \bar{g} + \epsilon \bar{\Omega} \cL_{\xi} \bar{g}  + \cO(\epsilon^2)  \\
& = \Omega \cL_{\chi} \bar{g} + \cO(\epsilon^2) \\
& = \Omega g + \cO(\epsilon^2) 
\end{align}
where $\Omega = \bar{\Omega} + \epsilon \cL_{\xi} \bar{\Omega}$.
It follows that if the background metric $\bar{g}_{\mu\nu}$ admits a conformal isometry generated by the conformal Killing vector $\bar{Y}^{\mu} \partial_{\mu}$, the perturbed metric generated by large-gauge transformations admits a deformed conformal isometry generated by the following conformal Killing vector
\begin{align}
Y^{\mu} \partial_{\mu} & = \left( \bar{Y}^{\mu} +\epsilon  [\xi, \bar{Y}]^{\mu} \right) \partial_{\mu} \\
& =  \left( \bar{Y}^{\mu} + \xi^{\alpha} \partial_{\alpha} \bar{Y}^{\mu} -  \bar{Y}^{\alpha} \partial_{\alpha} \xi^{\mu}\right) \partial_{\mu} \, .
\end{align}
Obviously, this applies to simple Killing vectors for which $\bar{\Omega} =0$. It follows that, if the seed background possesses enough Killing symmetries to integrate the geodesic motion, then the same is true for the new perturbative solution constructed by such soft symmetry, however complicated this new perturbative solution may be.

At this stage, one can wonder how the algebra of the background (conformal) Killing vectors transforms. To see this, let us consider two given vector fields of the background, \ie $\bar{X}^{\mu} \partial_{\mu}$ and $\bar{Y}^{\mu} \partial_{\mu}$, which satisfy
\be
[\bar{X}, \bar{Y}] = \bar{Z}\, .
\ee
This algebra is preserved under the action of the soft diffeomorphism. Indeed, consider the two Lie-dragged vectors $X = \cL_{\chi} \bar{X}$ and $Y = \cL_{\chi} \bar{Y}$, their algebra reads
\begin{align}
[X, Y] & = [\bar{X}, \bar{Y}] + \epsilon [\cL_{\xi} \bar{X}, \bar{Y}] + \epsilon [\bar{X} , \cL_{\xi} \bar{Y}] + \cO(\epsilon^2) \\
& = \bar{Z} + \epsilon \left(  [[\xi, \bar{X}], \bar{Y}] + [\bar{X}, [\xi, \bar{Y}]] \right) + \cO(\epsilon^2) \\
& =  \bar{Z} + \epsilon \left(  [[\xi, \bar{X}], \bar{Y}] + [[\xi, \bar{Y}], \bar{X}] \right) + \cO(\epsilon^2) \\
& = \bar{Z} + \epsilon [\xi, [\bar{X}, \bar{Y}]] + \cO(\epsilon^2) \\
& = \bar{Z} + \epsilon \cL_{\xi} \bar{Z} + \cO(\epsilon^2) \\
& = Z + \cO(\epsilon^2) 
\end{align}
where $Z = \cL_{\chi} \bar{Z}=  \bar{Z} + \epsilon \cL_{\xi} \bar{Z}$, which concludes the proof.

\bibliographystyle{JHEP}
\bibliography{biblio} 
\end{document}